\documentclass{aa}  
\usepackage{txfonts,graphicx}
\usepackage{longtable}
\usepackage{natbib}
\bibpunct{(}{)}{;}{a}{}{,}

\begin{document}

\title{Spectral properties of X-ray bright variable sources in the Taurus
Molecular Cloud}

\author{E. Franciosini\inst{1}
    \and  I. Pillitteri\inst{1,2}
    \and  B. Stelzer\inst{1}
    \and  G. Micela\inst{1}
    \and  K. R. Briggs\inst{3}
    \and  L. Scelsi\inst{2}
    \and  A. Telleschi\inst{3}
    \and  M. Audard\inst{4}\fnmsep\thanks{%
         \emph{Present address:} Integral Science Data Centre,
         Ch. d'Ecogia 16, 1290 Versoix, Switzerland 
	 \& Geneva Observatory, Ch. des Maillettes 51, 1290 Sauverny,
	Switzerland}
    \and  F. Palla\inst{5}
    \and  M. G\"udel\inst{3}
}

\offprints{E. Franciosini}

\institute{
  INAF - Osservatorio Astronomico di Palermo, Piazza del Parlamento 1, 90134
  Palermo, Italy\\
  \email{francio@astropa.unipa.it}
\and
  Dipartimento di Scienze Fisiche ed Astronomiche, 
  Universit\`a di Palermo, Piazza del Parlamento 1, 90134 Palermo, Italy
\and
   Paul Scherrer Institute, W{\"u}renlingen and Villigen, 5232 Villigen PSI,
   Switzerland 
\and
   Columbia Astrophysics Laboratory, Mail Code 5247, 550 West 120th Street,
   New York, NY 10027, USA
\and
   INAF - Osservatorio Astrofisico di Arcetri, Largo E. Fermi 5, 50125
   Firenze, Italy
}

\date{Received 10 October 2006 / Accepted 12 November 2006}

\abstract
{}
{We analyze 19 bright variable X-ray sources detected in the
\emph{XMM-Newton Extended Survey of the Taurus Molecular Cloud} (XEST), in
order to characterize the variations with time of their coronal properties
and to derive informations on the X-ray emitting structures.}
{We performed time-resolved spectroscopy of the EPIC PN and MOS spectra of
the XEST sources, using a model with one or two thermal components, and we
used the time evolution of the temperatures and emission measures during the
decay phase of flares to derive the size of the flaring loops.}
{The light curves of the selected sources show different types of
variability: flares, long-lasting decay or rise through the whole
observation, slow modulation or complex flare-like variability. Spectral
analysis shows typical quiescent plasma temperatures of $\sim$\,5\,--\,10~MK
and $\sim$\,15\,--\,35~MK; the cool component generally remains constant,
while the observed flux changes are due to variations of the hot component.
During flares the plasma reaches temperatures up to 100~MK and luminosities
up to $\sim 10^{31}$~erg~s$^{-1}$. Loop sizes inferred from flare analysis
are generally smaller than or comparable to the stellar radius.}
{}


\titlerunning{X-ray bright variable sources in the Taurus Molecular Cloud}

\maketitle


\section{Introduction}
\label{intro}

Young pre-main sequence (PMS) stars are well known to exhibit strong X-ray
emission at levels orders of magnitude higher than generally observed in
older active stars \citep{Feigel99,FavataMicela03,Guedel04}. Furthermore,
X-ray time variability is almost always detected in PMS stars, very often in
the form of flares, with a shape similar to solar flares, or of slow
modulation due e.g. to rotation \citep{Feigel99,Wolk05,Flaccomio05}. On the
Sun, the X-ray time variability is due to flare activity on timescales of a
few hours, to the appeareance of new active regions combined with the
surface rotation (timescales of a few weeks) and to the 11-yr solar activity
cycle. The flare activity observed in PMS stars suggests that a scaled
version of the solar corona should be present in these objects. The plasma
magnetically confined in loop-like structures would be heated by mechanisms
similar to those present in the solar corona, and would release the energy
in the form of flares. However, observations show that Classical T~Tauri
stars (CTTS), characterized by accretion from the circumstellar disk, are
statistically X-ray under-luminous with respect to Weak-lined T~Tauri stars
(WTTS), where disks are absent or weak
\citep{Stelzer01,Flaccomio03,Preibisch05,Francio06,Telleschi06}. Moreover,
high-resolution spectroscopy of the CTTS TW~Hya \citep{Kastner02,Stelzer04}
and BP~Tau \citep{Schmitt05} indicate the presence of very high densities
($n_\mathrm{e} \ga 10^{12}$~cm$^{-3}$) and a dominant cool plasma component
at 3~MK in TW~Hya, that have been attributed to emission from plasma heated
by an accretion shock. Enhanced X-ray emission and/or significant spectral
variations have been observed during accretion outbursts in V1647~Ori and
V1118~Ori \citep{Kastner04,Kastner06,Grosso04,Audard05}. These results
suggest that accretion may play a role in the X-ray emission process, either
by influencing the magnetic structure, or by providing an alternative X-ray
production mechanism.

The study of the time variability of the X-ray emission from PMS stars
allows us to gain insights into the structure and the heating mechanisms of
stellar coronae. In particular, the analysis of flares constitutes a
diagnostic tool to infer the size of the X-ray emitting structures.
Recently, \citet{Favata05} studied a sample of intense flares observed in
the Orion Nebula Cluster as part of the {\it Chandra} Orion Ultradeep
Project (COUP), finding that flares on PMS stars occur both in small loops,
of size less than a stellar radius, similar to older active stars, and in
large loops, up to $10-20\,R_\star$, likely connecting the stellar surface
and the circumstellar disk.

In this paper we study the X-ray properties of a sample of PMS stars of the
Taurus Molecular Cloud (TMC) showing significant time variability. The TMC
is one of the nearest regions of star formation ($d = 140$~pc), containing
$\sim$\,340 known members in an area $\sim$\,100 square degrees large, and
is characterized by a low stellar density (1\,--\,10~pc$^{-3}$), by the lack
of massive stars and by a significant fraction of binary or multiple
systems. Star formation appears to have occurred in several epochs during
the last 10~Myr. 

Our work complements the study by \citet{Stelzer06}, who analyze the X-ray
time variability of TMC sources from a statistical point of view, and derive
the frequency and energy distribution of flaring events. In our paper we
concentrate on the time-dependent spectral analysis of the brightest
sources, showing both flares and other kinds of variability, in order to
investigate the changes of the plasma parameters and to derive information
on the X-ray emitting structures.

The paper is organized as follows. In Sect.~\ref{selection} we describe the
target selection and the data analysis. In Sect.~\ref{results} we present
the results obtained for TMC members, whereas in Sect. \ref{nonmembers} we
present the variability of three sources not related to the TMC. Discussion
and conclusions are given in Sect.~\ref{concl}.


\section{Source selection and analysis}
\label{selection}

\subsection{Observations}
\label{obs}

Our study is based on the data obtained from the \emph{XMM-Newton Extended
Survey of the Taurus Molecular Cloud} (XEST), a wide X-ray survey performed
with the XMM-{\it Newton} satellite aimed at studying the properties of
young PMS stars in the TMC \citep{Guedel06}. The survey consists of 19
fields of $\sim$\,33~ks exposure each, covering the densest regions of the
cloud, plus 9 archival exposures lasting up to $\sim$\,130~ks; the total
area covered by the survey is about 5 square degrees.

A detailed description of the survey, including details on the primary
reduction of the raw EPIC MOS and PN datasets, on the source detection
procedure and on the definition of the extraction regions for the source and
background light curves and spectra is reported in \citet{Guedel06}.

The work presented here uses data from all XEST fields except for XEST-01
and XEST-25 that are dedicated to separate projects \citep[][Grosso et al.,
in preparation]{Guedel06tt}. We also excluded the classical T~Tauri star
SU~Aur (source XEST-26-067) that is discussed in detail in a separate paper
\citep{Francio07su}. We also note that two of the fields, namely XEST-23 and
XEST-24, correspond to two consecutive observations with the same pointing
position, with a duration of $\sim$\,70 and 40~ks, respectively, separated
by $\sim$\,5~ks. Sources detected in both exposures have been treated as a
single source in the following analysis, with the two exposures added
together.

\begin{table*}
\centering
\caption{Properties of the sample of XEST variable sources analyzed in this
paper. Stellar parameters are taken from \citet{Guedel06}. In the last
column we give a classification of the observed variability.} 
\label{tab_sample}
\begin{tabular}{llllll} 
\hline\hline\noalign{\smallskip}
XEST ID      & Optical ID  & Sp. T.& TTS class.& $R_\star/R_\odot$& 
Variab. type\\
\noalign{\smallskip}\hline \noalign{\smallskip}  
04-016       & V830 Tau    & K7     & WTTS     & 1.79  & slow decay      \\
09-026       & HQ Tau AB   & \ldots & WTTS     & \ldots& slow decay      \\
11-057       & FS Tau AC   & M0+M3.5& CTTS+CTTS& 0.93  & slow decay      \\
12-040       & DN Tau      & M0     & CTTS     & 2.25  & slow modulation \\
15-040       & DH Tau AB   & M1     & CTTS     & 1.82  & slow decay      \\
17-066       & JH 108      & M1     & WTTS     & 1.32  & atypical flare  \\
21-039       & HD 283572   & G5     & WTTS     & 2.56  & slow modulation \\
22-047       & XZ Tau AB   & M2+M3.5& CTTS+CTTS& 1.18  & slow rise       \\
22-089       & L1551 51    & K7     & WTTS     & 1.39  & slow decay      \\
23-032/24-028& V410 Tau ABC& K4     & WTTS     & 2.31  & atypical flare  \\
23-047/24-040& V892 Tau    & B9     & Herbig Ae& 2.66  & atypical flare  \\
23-050/24-042& V410 X7     & M0.75  & WTTS     & 1.69  & flares\\
23-056/24-047& Hubble 4    & K7     & WTTS     & 3.33  & complex variab. \\
23-074/24-061& V819 Tau AB & K7     & WTTS     & 1.93  & flare \\
26-072       & HBC 427     & K7     & WTTS     & 1.85  & flare \\
28-100       & BP Tau      & K7     & CTTS     & 1.97  & flare \\
\noalign{\smallskip}\hline\noalign{\smallskip}
05-031       & HD 283810   & K5V    &          &       & flare \\
16-031       & 2M J04195676+2714488& \ldots &  &       & flares\\
22-024       & HD 285845   & G6     &          &       & complex variab. \\
\noalign{\smallskip}\hline
\end{tabular}
\end{table*}

\subsection{Analysis of light curves and sample selection}
\label{lc}

To select the sample of sources for this study, we first extracted, for all
detected X-ray sources, the PN and MOS source and background photons in the
energy band 0.3\,--\,7.3~keV, using the extraction regions defined by
\citet{Guedel06}. In order to have a continuous time coverage for each
observation, without gaps in the light curves, we chose not to filter out
the periods of high background count rate due to proton flares. We then
limited our analysis to sources with at least 1500 net counts in the PN
exposures (or in the MOS exposures in the cases where the PN was not
available), in order to have sufficiently high count statistics for good
time-resolved spectral analysis. This selection necessarily biases our study
towards the X-ray brightest and likely most active sources.

For these sources, we extracted the PN (or MOS) light curves, and applied to
them the \emph{Maximum Likelihood Blocks} (MLB) method described by
\citet{Stelzer06}, in order to identify variable sources and to define the
intervals to be used for time-dependent spectroscopy. The MLB algorithm
divides the full exposure into time intervals (blocks) where the source is
assumed to be constant; the boundaries of the time blocks are chosen such
that between two consecutive blocks the mean count rate changes by more than
a given significance threshold. The significance threshold for the change
points, at the 99\% level, was determined through simulations of constant
rate light curves (Flaccomio et al., in preparation). We set a minimum
number of 750 net source photons for each block, in order to have enough
counts in each spectrum to perform a reliable spectral analysis. We note
that this choice does not allow us to examine rapid variations,
small-amplitude flaring events and low-level variability.

The variability of the background in many XEST fields was significant and
was taken into account using the following procedure: the MLB algorithm was
used to split the background light curve into blocks of constant count rate
level; for each of these blocks, the background level was scaled to the
source extraction area to yield the number of expected background photons in
the source area for the given time interval. The resulting number of photons
is then removed from the source event file uniformly across each background
block. The result is a ``source-only" event file, i.e. a
background-subtracted photon time series for the source.
 
Using the MLB results, we restricted our sample by considering only those
sources whose light curves showed significant variability, i.e. those that
have been divided by the algorithm in at least two blocks with mean count
rates differing by more than $3\sigma$. This selection process led to a
sample of 16 variable X-ray sources associated with known PMS members of the
TMC, plus additional three sources not related to the region or that are not
confirmed members of the TMC. The weakest source in the final sample is
JH~108, with $\sim$\,2500 net PN counts.  The main properties of the
selected sources are given in Table~\ref{tab_sample}. The three non-members
will be discussed separately in Sect.~\ref{nonmembers}. 

The 16 TMC members are mostly late-K and early-M stars; five of them are
classified as CTTS, ten as WTTS, and one is a Herbig~Ae star. Six of these
sources (including the Herbig~Ae star) are binary or multiple systems
unresolved in X-rays with XMM-{\it Newton}.

\subsection{Time-resolved spectroscopy}
\label{spectroscopy}

We have performed a time-dependent spectral analysis of the selected
sources, in order to study the changes with time of the plasma
characteristics, i.e. temperature and emission measure (EM), that give us
information on the origin of the emission, and to derive, in the case of
flares, information on the size of the flaring region from flare modeling. 

PN and MOS spectra for each source and each time block have been extracted
from the source and background event files. For the response matrices, we
used the appropriate canned response matrix files for PN and MOS, and the
ancillary response files produced for each individual source \citep[see][
for details]{Guedel06}. The PN and MOS spectra have been rebinned in order
to have at least 15 counts per bin, and have been jointly fitted in XSPEC
v.11.3.2, using a thermal VAPEC model with one or two temperature
components, plus a common photoelectric absorption component. The hydrogen
column density $N_\mathrm{H}$ was left as a free parameter, except in a few
cases where it was not constrained by the fit, and it was therefore fixed to
a value derived for other blocks of the same source. Abundances were kept
fixed, with their values following a pattern derived from estimates of
coronal abundances of X-ray active young stars\footnote{
The abundances used, relative to the solar abundances by \citet{AG89}, are:
C\,=\,0.45, N\,=\,0.788, O\,=\,0.426, Ne\,=\,0.832, Mg\,=\,0.263,
Al\,=\,0.5, Si\,=\,0.309, S\,=\,0.417, Ar\,=\,0.55, Ca\,=\,0.195,
Fe\,=\,0.195, Ni\,=\,0.195.} 
\citep{Scelsi05,Telleschi05,Argi04}. The best-fit results are reported in
Table~\ref{fit}; errors are 90\% confidence intervals for one interesting
parameter.

\begin{figure*}
\centering
\includegraphics[width=\hsize]{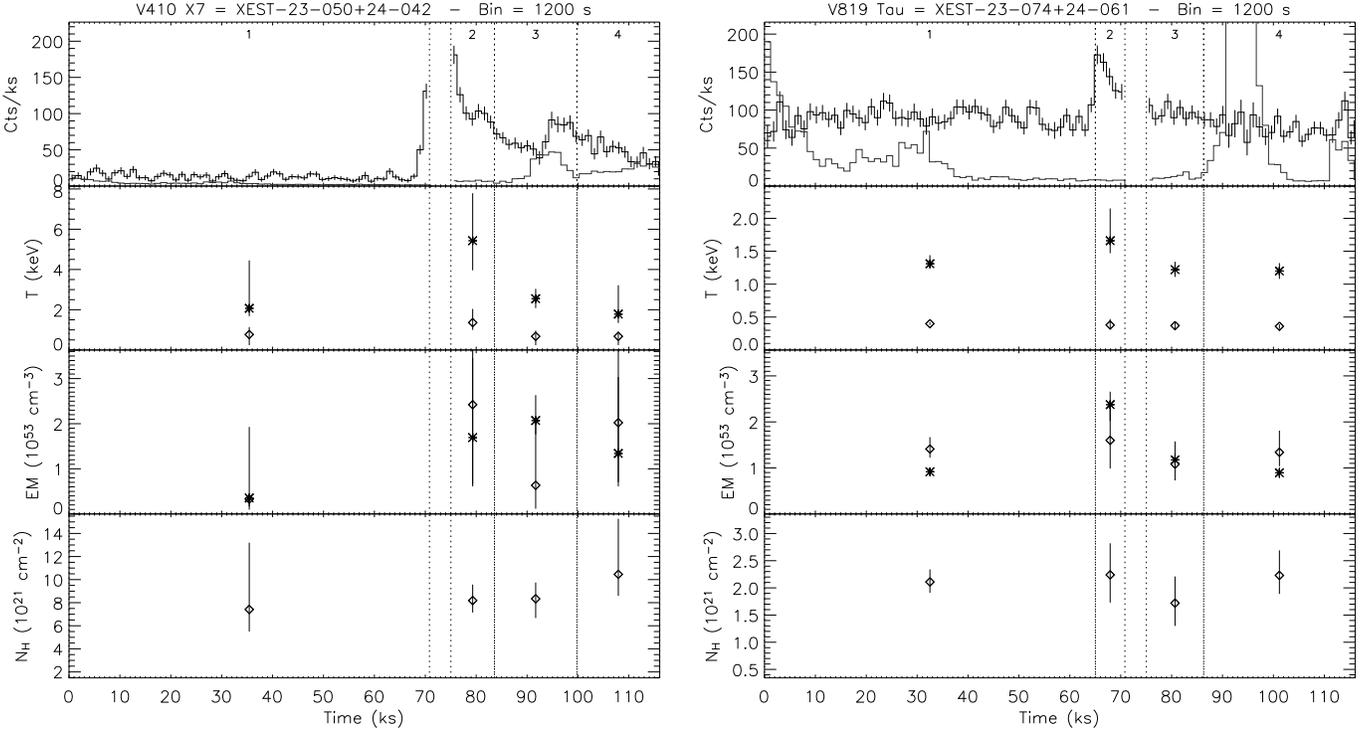}
\caption{Light curves and spectral fitting results for the flaring TMC
sources V410~X7 (XEST-23-050/24-042) and V819~Tau (XEST-23-074/24-061). In
the \emph{top panels}, the background-subtracted PN light curve of each
source is plotted with a thick solid line, while the thin solid line shows
the background light curve, scaled to the source area. The vertical dotted
lines mark the time intervals, derived from the MLB analysis, used for the
time-resolved spectroscopy. In the \emph{other panels} we show, from top to
bottom, the time evolution of the best-fit temperatures, emission measures
and hydrogen column density. The temperatures and EMs of the cool and hot
component are indicated by diamonds and asterisks, respectively.}
\label{impfl} 
\end{figure*}

\begin{figure*}
\centering
\includegraphics[width=\hsize]{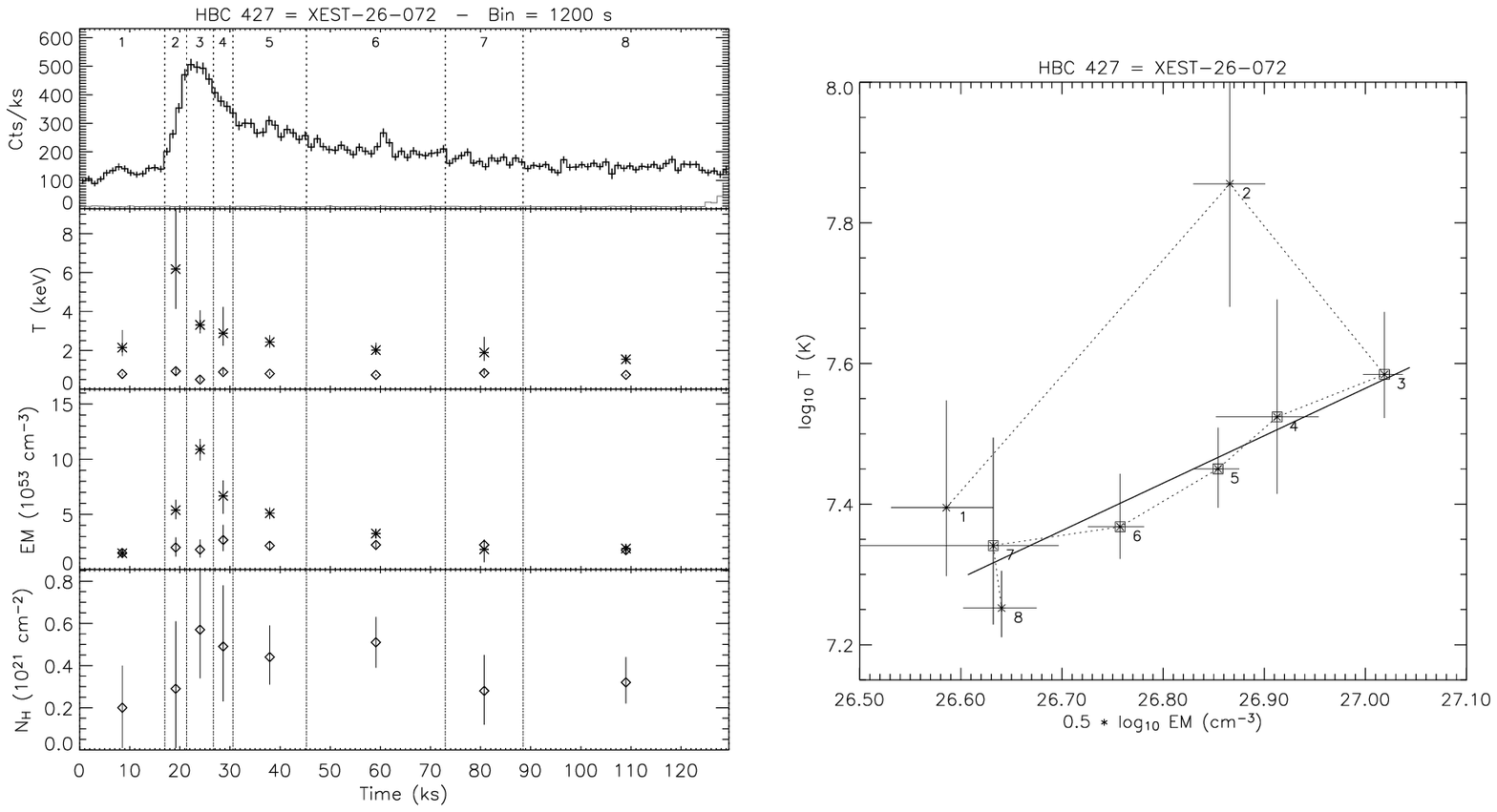}
\caption{{\em Left}: Light curve and spectral fitting results for HBC~427
(XEST-26-072). Panels are the same as in Fig.~\ref{impfl}. Note that in this
case we show the combined MOS1+MOS2 light curve, the PN being not available.
{\em Right}: Evolution of the best-fit temperature and EM of the hot
component for HBC~427. To better clarify the trend with time of the
parameters, the points are connected by dotted lines. The thick solid line
shows the linear fitting to points \#3 to \#7, corresponding to the flare
decay; for clarity they are marked with open squares.} 
\label{xest26072}
\end{figure*}

Figs.~\ref{impfl}-\ref{complex} show the light curves and spectral
parameters derived for TMC sources; the results for TMC non-members are
shown in Figs.~\ref{nonm}-\ref{xest16031}. For each source, we plot in the
top panel the PN (or MOS) background-subtracted light curve and the
background light curve, while in the other panels we show the evolution of
the best-fit temperatures and EMs and of the hydrogen column density. The
time intervals derived from the MLB method and used for the analysis are
indicated by dotted vertical lines. In Figs.~\ref{xest26072},
\ref{xest28100}, \ref{xest23032} and \ref{xest16031} we also show, in the
right panels, the evolution of the plasma temperature and emission measure
used for flare analysis.

\subsection{Flare modeling}
\label{modeling}

Flare light curves contain information on heating and cooling parameters and
indirectly on the density and geometry of the flaring sources. As stellar
flare observations are not spatially resolved, various techniques have been
developed to use the light curves and measurable quantities such as the
emission measure or the electron temperature to deduce flare source
parameters \citep[see review by][]{Guedel04}. Each of these methods
introduces simplifications and makes various assumptions on the heating and
cooling processes, often adopted by analogy with solar flares. A
particularly straightforward physical model has been developed by
\citet{Reale97}, that derives the heating fraction and the magnetic loop
length from hydrodynamic considerations, using only the run of flare
temperature and emission measure. Its advantage is its simple application to
stellar observations; it has also been tested on moderate single-loop solar
flares. It is, on the other hand, not applicable to flares occurring in
magnetic arcades as in gradual solar flares, where a large number of
magnetic structures is ignited and cooling in sequence. Then, both the rise
and the decay time are essentially determined by the history of the heating
energy release rather than by cooling physics
\citep[e.g.][]{Pneuman82,KP84}. Comparative studies have been presented in
\citet{Guedel04prox} and \citet{Reale04} for an exceptionally strong flare
on the nearby active star Proxima Centauri. We are not in a position to
characterize flares on T~Tau stars here. We henceforth will adopt the model
of \citet{Reale97} to discuss possible systematics in our results.

In the case of a single flaring loop decaying freely after an initial
impulsive heating, \citet{Serio91} showed that the loop semi-length $L$ is
related to the thermodynamic decay time $\tau_\mathrm{th}$ by the relation:
\begin{equation}
L = \frac{\tau_\mathrm{th} \sqrt{T_\mathrm{max}}}{3.7 \times
10^{-4}} \; ,
\label{l_free}
\end{equation}
where $T_\mathrm{max}$ is the maximum temperature of the flaring plasma.
However, if significant heating is present also during the decay phase, the
flare decay time will be longer than $\tau_\mathrm{th}$, and the above
equation would lead to an overstimate of the loop length. Using
hydrodynamical simulations of flaring loops taking into account the presence
of prolonged heating, \citet{Reale97} showed that the slope $\zeta$ of the
flare decay path in a diagram of $\log T$ vs $\log n_\mathrm{e}$ (or
equivalently $\log\sqrt{EM}$) is a diagnostic of the presence of residual
heating during the flare decay. The loop semi-length $L$ can then be derived
from $\zeta$ and the observed flare decay timescale $\tau_\mathrm{lc}$ using
the following relation:
\begin{equation}
L = \frac{\tau_\mathrm{lc} \sqrt{T_\mathrm{max}}}{3.7 \times 10^{-4}\;
F(\zeta)} \; ,
\label{l_loop}
\end{equation}
where $F(\zeta)$ is the ratio between the observed and thermodynamic decay
times, which depends on the amount of residual heating. $T_\mathrm{max}$, as
in Eq.~(\ref{l_free}), is the maximum temperature in the flaring loop, that
can be derived from the flare peak temperature $T_\mathrm{obs}$ obtained
from the spectral fits. The expression of $F(\zeta)$ and the relationship
between $T_\mathrm{max}$ and $T_\mathrm{obs}$ depend on the instrumental
response. In the case of EPIC PN we have \citep{Giardino06}:
\begin{equation}
T_\mathrm{max} = 0.13 \; T_\mathrm{obs}^{1.16} \;.
\label{tmax}
\end{equation}
and
\begin{equation}
F(\zeta) = 1.36 + \frac{0.51}{\zeta - 0.35}
\label{fzeta}
\end{equation}
The expression of $F(\zeta)$ is valid only for $0.35<\zeta\le 1.6$, where
the lower limit corresponds to the case where the heating timescale is
comparable to the flare decay timescale, while the upper limit corresponds
to a freely-decaying loop with no residual heating after the impulsive
energy release.

The above formulae are calibrated on the EPIC PN response, however they give
a correct order of magnitude estimate of the loop length also in the case of
MOS, given the similarity of the instrumental responses and the wide
spectral band used. We have therefore applied them also to the analysis of
the flare of HBC~427, for which PN is not available.

\section{Sources associated with TMC members}
\label{results}

We have divided the sources associated with TMC members into three classes
according to the type of variability observed in their light curves: {\em
(i)} sources showing evident flares, either with a ``classical'' shape, with
fast rise and slower decay, or with a gradual rise and/or a symmetrical
shape; {\em (ii)} sources with slow prolonged decay or rise of the count
rate; and {\em (iii)} sources with slow modulation or complex variability
that could be due to superimposed flaring events.

\subsection{Flares}
\label{fl} 

\begin{figure*}
\centering
\includegraphics[width=\hsize]{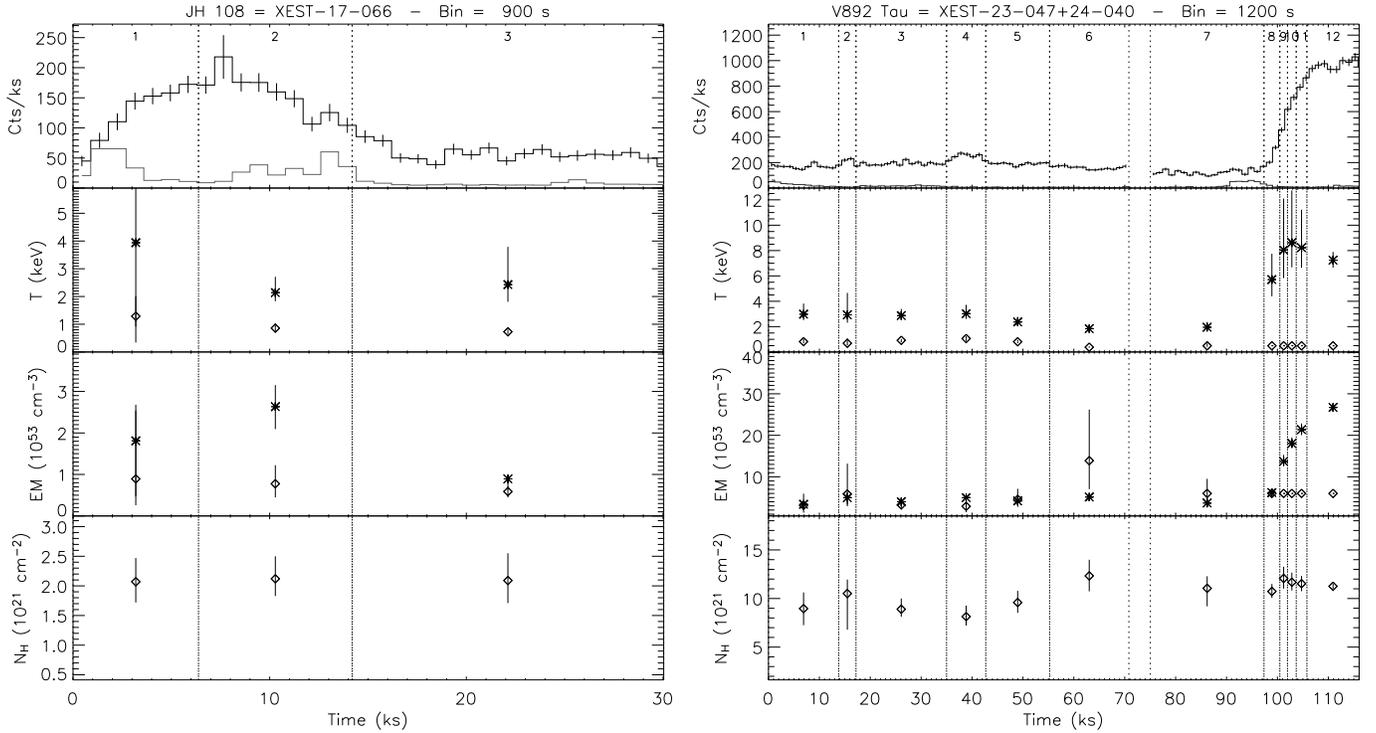}
\caption{Light curve and spectral fitting results for the TMC sources JH~108
(XEST-17-066) and V892~Tau (XEST-23-047/24-040) showing gradual rise flares.
Panels are the same as in Fig.~\ref{impfl}.}
\label{atypfl} 
\end{figure*}

Seven of the TMC sources show evident flares in their light curve. Four of
them (V410~X7, V819~Tau, HBC~427 and BP~Tau) show the typical flare
behaviour, characterized by a rapid increase of the count rate and of the
plasma temperature, followed by a slower decay. Their light curves and the
time evolution of the plasma parameters are shown in
Figs.~\ref{impfl}-\ref{xest28100}. The other three sources (JH~108, V892~Tau
and V410~Tau) show peculiar flaring variability, i.e. flares that are
characterized by a gradual rise phase with a flat top and/or a nearly
symmetrical shape. Their light curves and evolution of the plasma parameters
are shown in Figs.~\ref{atypfl} and \ref{xest23032}.

\subsubsection{V410~X7 (XEST-23-050/24-042)}

The light curve of \object{V410~X7} (XEST-23-050/24-042, M0.75, WTTS) shows
two flares that occurred after an initial quiescent period lasting for
$\sim$\,65~ks. Unfortunately, the XEST-23 exposure stopped just after the
start of the rise phase of the first flare, while at the beginning of the
XEST-24 exposure the flare was already decaying; therefore the flare peak
was not observed. The flare is very strong, with an increase of the count
rate by more than an order of magnitude. The decay phase is initially rapid,
with an e-folding timescale of $\sim 2.1$~ks; then a slower decay is
observed, after a small bump occurring about 5~ks after the start of the
XEST-24 exposure. A second smaller flare, with a decay e-folding time of
$\sim$\,20~ks, is superimposed on the late decay phase. As mentioned in
Sect.~\ref{lc}, our choice of the block size does not allow us to derive the
plasma characteristics of this second flare separately from the decay of the
first one.

The quiescent emission has temperatures of $\sim$\,9 and 24~MK with similar
EMs. The first flare reached a temperature $>60$~MK at the peak, as
estimated from the best-fit temperature of block \#2 where the flare is
already decaying. Taking this value as peak temperature and the decay time
of 2.1~ks, and applying to this flare the formula for a freely-decaying loop
(Eq.~(\ref{l_free})), we can put a lower limit on the loop semi-length of
$L\ge 5 \times 10^{10}$~cm, corresponding to $\sim 0.4\,R_\star$.

\subsubsection{V819~Tau (XEST-23-074/24-061)}

\object{V819~Tau} (XEST-23-074/24-061) is a binary WTTS with spectral type
K7. It showed a small impulsive flare at the end of the XEST-23 exposure,
with a factor of 2 increase in the count rate, whose decay continues in the
XEST-24 observation. The decay phase has an e-folding time of $\sim$\,10~ks.
The flare is not very hot, with $T_2 =$\,20~MK in block \#2, not much higher
than the quiescent value of $\sim$\,15~MK, although the true peak
temperature is likely hotter since this time interval includes also a
significant part of the decay phase. We also note that the cool component
remained constant, within the errors, for the whole observation. We can
obtain a rough estimate of the loop semi-length using the temperatures and
EMs of the hot component in blocks \#2 and \#3, from which we derive
$\zeta\sim\,0.9$ and $L\sim 7 \times 10^{10}$~cm, corresponding to
$0.5\,R_\star$.

\begin{figure*}
\centering
\includegraphics[width=\hsize]{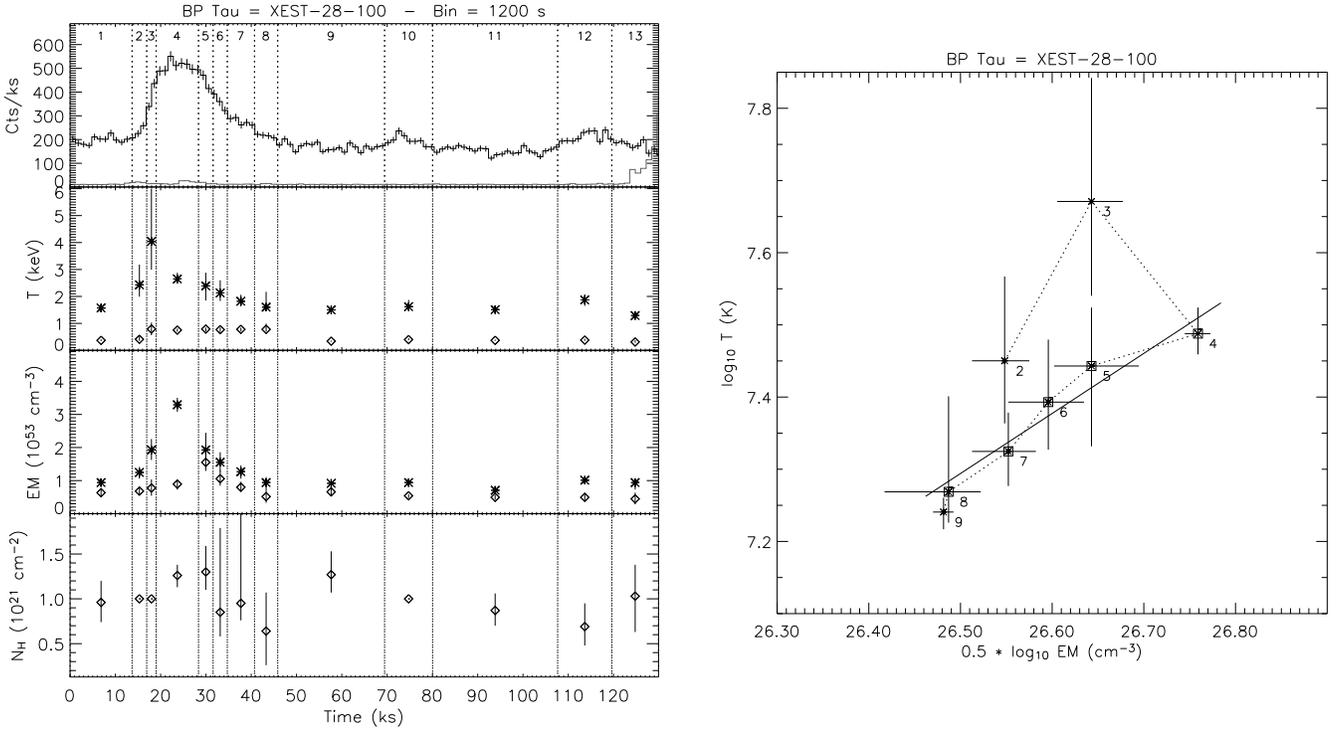}
\caption{{\em Left}: Light curve and spectral fitting results for BP~Tau
(XEST-28-100). Panels are the same as in Fig.~\ref{impfl}. {\em Right}:
Evolution of the best-fit temperature and EM of the hot component. See
Fig.~\ref{xest26072} for details. For the sake of clarity, we plot only the
points from \#2 to \#9. The slope has been fitted between points \#4 and
\#8.} 
\label{xest28100}
\end{figure*}

\subsubsection{HBC~427 (XEST-26-072)}

During the XMM-{\it Newton} observation, the K7 WTTS \object{HBC~427}
(XEST-26-072) underwent a strong long-duration flare that covered nearly the
entire exposure time of $\sim$\,130~ks. After a short initial quiescent
phase, the count rate increased by a factor of 5 in $\sim$\,6~ks, then it
decreased slowly, returning to a level similar to the preflare one after
$\sim\,70$~ks. The decay e-folding time is $20.2 \pm 1.0$~ks. The time
evolution of the plasma parameters follows the typical behaviour observed in
flares, with the temperature peaking during the rise phase and the emission
measure peaking at the flare peak. The quiescent emission has temperatures
of $\sim$\,8 and 25~MK with equal EMs. While the temperature and EM of the
coolest component does not change significantly during the flare, the hot
component reaches a temperature as high as $\sim$\,72~MK; therefore, for the
flare analysis we will assume that the flaring plasma is described by the
hot component.
 
In the right panel of Fig.~\ref{xest26072} we show the evolution of the
flare in the $T$ vs $\sqrt{EM}$ plane. The maximum loop temperature, derived
from Eq.~(\ref{tmax}), is $\sim$\,168~MK. From the linear fitting of the
points from block \#3 to block \#7 we find a slope $\zeta = 0.67 \pm 0.09$.
This value indicates the presence of significant residual heating after the
initial ignition. Combining the observed decay time with the above value of
$\zeta$, from Eq.~(\ref{l_loop}) we derive $L = 2.4 \pm 0.4 \times
10^{11}$~cm, i.e. $\sim$\,2 stellar radii.

From the EM of the hot component in block \#2 we can estimate the mean
electron density in the loop at the peak of the flare, assuming a loop with
typical aspect ratio $R_\mathrm{loop}/L = 0.1$, as commonly observed in
solar flares: we find $n_\mathrm{e}\sim 2 \times 10^{10}$~cm$^{-3}$,
consistent with the values found for the COUP flares with similar loop size
\citep{Favata05}.

\subsubsection{BP~Tau (XEST-28-100)}

\object{BP~Tau} (XEST-28-100) is a K7 CTTS. Its light curve shows a flare
beginning $\sim$\,15~ks after the start of the observation, with an increase
of the count rate by a factor of $\sim$\,2.5 in $\sim$\,5~ks and a total
duration of $\sim$\,35~ks. The e-folding decay time is $10.4 \pm 0.7$~ks.
After the flare the X-ray emission shows low-level variability until the end
of the observation. As for HBC~427, BP~Tau shows the typical flare behaviour
of the plasma parameters, with the temperature peaking in the rise phase
before the EM. The quiescent emission before the flare has temperatures of
$\sim$\,4 and 18~MK with $EM_2/EM_1 =1.5$; the temperature of the cooler
component increases to $\sim$\,9~MK during the flare, while the hotter
component reaches $\sim$\,50~MK. Both EMs increase significantly during the
flare.
 
The right panel of Fig.~\ref{xest28100} shows the evolution of the flare
temperature and EM. From Eq.~(\ref{tmax}), we derive a maximum loop
temperature of $\sim$\,100~MK. The slope between points \#4 and \#8 is
$\zeta = 0.83 \pm 0.13$, indicating that residual heating is present during
the decay. Using Eq.~(\ref{l_loop}) we derive $L = 1.2 \pm 0.1 \times
10^{11}$~cm, comparable to the stellar radius. The estimated density at the
flare peak is $n_\mathrm{e}\sim 3 \times 10^{10}$~cm$^{-3}$, similar to the
value found for HBC~427.

\subsubsection{JH\,108 (XEST-17-066)}

\object{JH\,108} (XEST-17-066, WTTS, spectral type M1) shows a symmetrical
flare, with a gradual increase of the count rate by a factor of $\sim$\,4 in
$\sim$\,8~ks, followed by a decay with similar duration; the decay e-folding
time is 3.3~ks. This source has quiescent temperatures of 8 and 30~MK and EM
ratio of $\sim$\,1.5. As before, the cool component remains nearly constant,
while for the hotter component the temperature is higher during the rise
phase, reaching 45~MK, and the EM peaks in the second time block,
corresponding to the peak and decay phases. A rough estimate using
Eq.~(\ref{l_free}) gives $L \la 6\times 10^{10}$~cm, i.e.
$\sim\,0.7\,R_\star$.

\subsubsection{V892~Tau (XEST-23-047/24-040)}

\object{V892~Tau} = Elias\,1 (XEST-23-047/24-040) is a triple system
composed of a Herbig~Ae star, a low-mass companion of spectral type
$\sim$\,M2 at $4.1\arcsec$ \citep[Elias\,1\,NE,][]{Leinert97} and a
recently-discovered close companion at $0.05\arcsec$ with a mass of $1.5-2
\,M_\odot$ \citep{Smith05}. 

For the first 95~ks of the observation this source showed low-level
variability, with temperatures in the range $T_1=5-10$~MK and
$T_2=20-35$~MK. Then the source underwent a strong flare, with a gradual
rise lasting for $\sim$\,7~ks and ending in a plateau, at a level a factor
of $\sim$\,10 higher than the pre-flare count rate, where the emission
remained nearly constant for at least 10~ks, until the end of the
observation. For the spectral analysis during the flare, the parameters of
the cool component were kept fixed since they were not constrained, and the
flare evolution was entirely dominated by the hot component. The plasma
heating is strong in the first block of the rising phase, where the
temperature increases from $\sim$\,25 to $\sim$\,65~MK; then the temperature
stays at a nearly constant level, within the errors, around
$\sim$\,95\,--\,100~MK in the following three blocks, and starts to decrease
only when the count rate reaches the plateu in the last time block. The
emission measure on the other hand continues to increase reaching its
maximum in the last block. The observed detailed evolution of the
temperature and EM is a clear example of the flare heating process, in
which, following the energy release, accelerated electrons precipitating
into the chromosphere rapidly heat the plasma, and produce an evaporation of
material which gradually fills the loop increasing the emission measure.

The long rise time suggests that a large coronal structure is involved in
the flare. This flare has been studied previously by \citet{Giardino04}, who
modeled the rise phase using detailed hydrodynamical simulations, finding a
loop size of the order of $1\times 10^{11}$~cm, equal to $\sim 0.5
\,R_\star$, and that a magnetic field in excess of 500~G is required to
confine the flaring plasma. Based on the position of the X-ray source and on
the comparison with a {\it Chandra} observation where the primary and
Elias\,1\,NE are resolved, \citet{Giardino04} concluded that the flare
occurred in the corona of the Herbig~Ae star, which would imply the
existence of a convective zone in the outer layers of the star, generally
not expected in stars of this spectral type. However, the discovery of a
lower-mass close companion suggests that the observed X-ray emission and the
flare more likely originate from it rather than the Herbig star itself. This
is supported by the high temperatures (20\,--\,30~MK) observed in this
source and in other Herbig~Ae/Be stars \citep{Stelzer06hae}, similar to
those commonly found for later-type PMS stars but significantly higher than
those ($\sim\,5-6$~MK) found in the Herbig stars HD~163296 and AB~Aur
\citep{Swartz05,Telleschi06ab}.

\begin{figure*}
\centering
\includegraphics[width=\hsize]{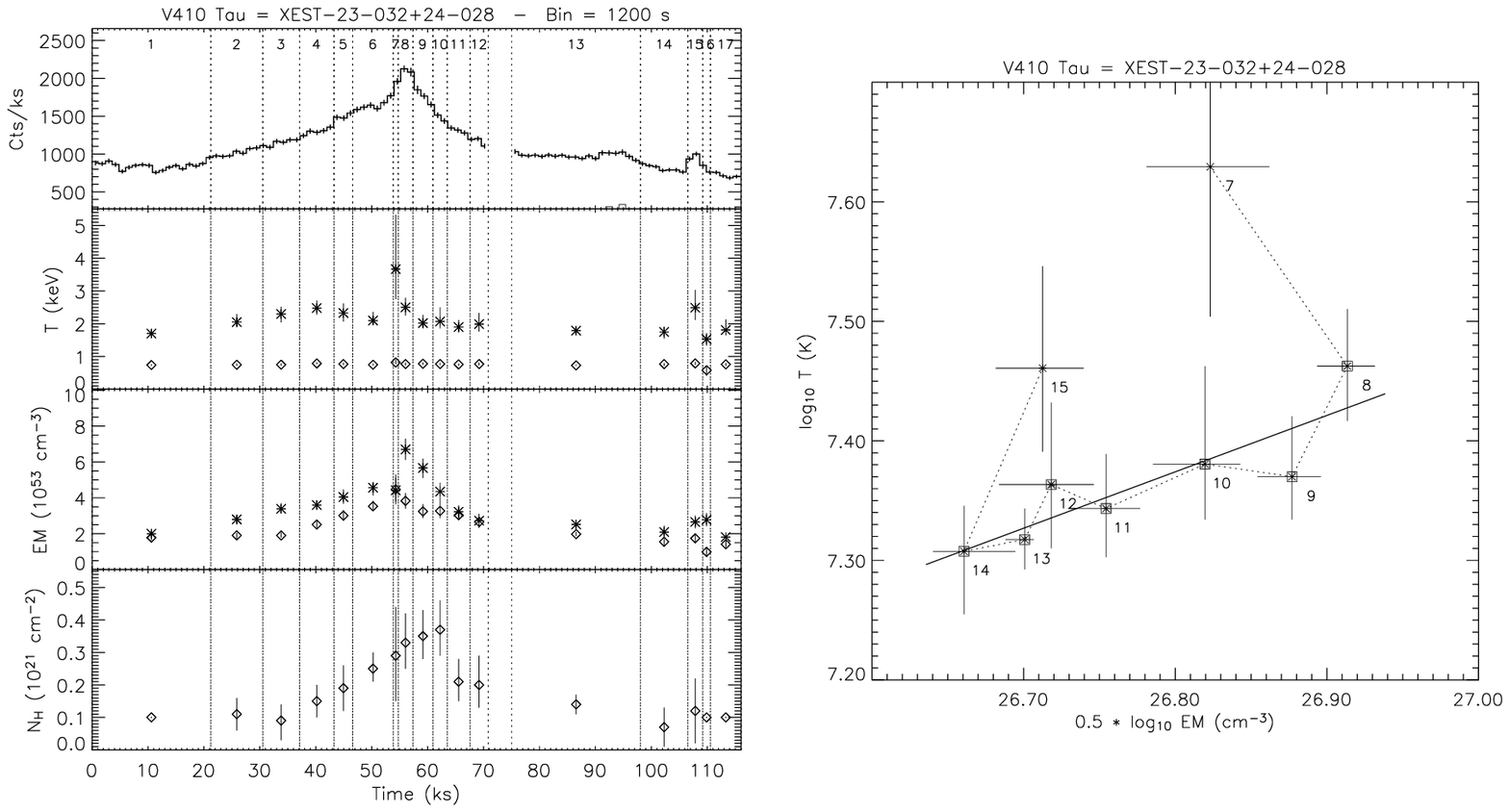}
\caption{{\em Left}: Light curve and spectral fitting results for V410~Tau
(XEST-23-032/24-028). Panels are the same as in Fig.~\ref{impfl}. {\em
Right}: Evolution of the best-fit temperature and EM of the hot component.
See Fig.~\ref{xest26072} for details. For the sake of clarity, we plot only
the points from \#7 to \#15. The slope has been fitted between points \#8
and \#14.}
\label{xest23032}
\end{figure*}

\subsubsection{V410~Tau (XEST-23-032/24-028)}

The WTTS triple system \object{V410~Tau} (XEST-23-032/24-028) shows a very
peculiar light curve with a nearly symmetrical triangular shape. The count
rate rises gradually during the first 55~ks of observation until block \#8,
where it reaches a peak a factor of $\sim 3$ higher than at the start of the
observation. Then the emission decays until the last block; a second, weaker
flare is superimposed on the decay $\sim$\,9~ks before the end of the
observation. Assuming that the first 18~ks of the observation represent the
quiescent level, we find that the rising phase has an e-folding time of
$\sim\,20\pm 1$~ks. After the peak the emission decays initially with an
e-folding time of $8.1\pm 0.4$~ks until the end of the XEST-23 exposure; at
the beginning of the XEST-24 exposure the emission stays at a steady level
for 23~ks (block \#13), then the decay restarts, and continues until the end
of the observation, interrupted only by the small flare. 

During the observation, the temperature of the cool component does not vary
significantly, and the flare evolution is determined by the variations of
$T_2$ and of the EMs. In the rise phase the hot temperature initially
increases from 20 to 30~MK, reaching a maximum in block \#4, then decreases
again in the following two time intervals; a sudden increase to
$\sim$\,45~MK occurs in block \#7, just before the peak. The EMs of both
plasma components increase during the entire rise phase, reaching their
maximum at the flare peak, as commonly observed in solar and stellar flares.
After the peak the temperature and the EMs decrease returning to the
quiescent value. A new temperature increase is observed in block \#15,
corresponding to the second small flare. The long rise time, combined with
the slow change of the plasma temperature during the rise phase, might
indicate that we are observing a rotationally modulated flare, or the
superposition of a flaring event and an underlying modulated emission.

\begin{figure*}
\centering
\includegraphics[width=\hsize]{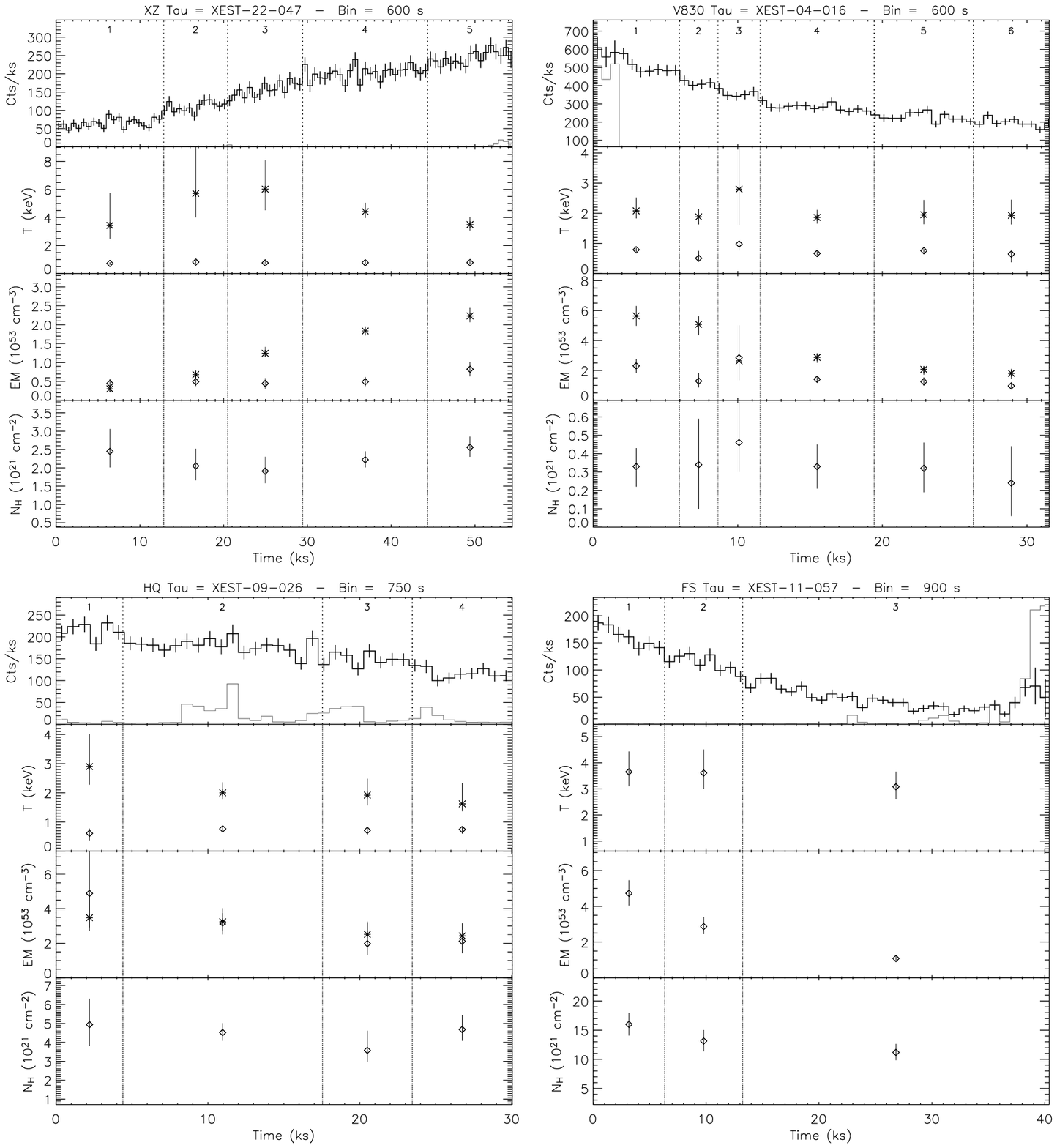}
\caption{Light curve and spectral fitting results for TMC sources showing a
long-lasting rise (XZ~Tau = XEST-22-047) or decay (V830~Tau = XEST-04-016,
HQ~Tau = XEST-09-026, FS~Tau = XEST-11-057). Panels are the same as in
Fig.~\ref{impfl}.}
\label{longdec1} 
\end{figure*}

We also note that the best-fit colum density increases by a factor of 4
during the flare, following a trend similar to that of the EMs. However, we
found equally acceptable fits, although with slightly higher values (by at
most 10\%) of the reduced $\chi^2$, by keeping $N_\mathrm{H}$ fixed to
$1\times 10^{20}$~cm$^{-2}$, as found in the first time intervals, without
significant changes in the time evolution of the other parameters. Since
this star is a WTTS, which is not expected to possess circumstellar material
that could justify changes in the intervening absorption, and that such low
values of $N_\mathrm{H}$ are at the sensitivity limit of EPIC, we are not
confident on the significance of the observed column density variations.

The right panel of Fig.~\ref{xest23032} shows the evolution of the
temperature and EM of the hot component during the flare. The slope during
the decay, between blocks \#8 and \#14, is $\zeta= 0.47 \pm 0.12$,
indicating that the decay of the flaring structure is largely determined by
the heating decay timescale. In this case the loop size is poorly
constrained, given the large uncertainty on $\zeta$: we obtain in fact $L
\sim 4 \pm 3 \times 10^{10}$~cm, or $0.1-0.4\,R_\star$.

\begin{figure*}
\centering
\includegraphics[width=\hsize]{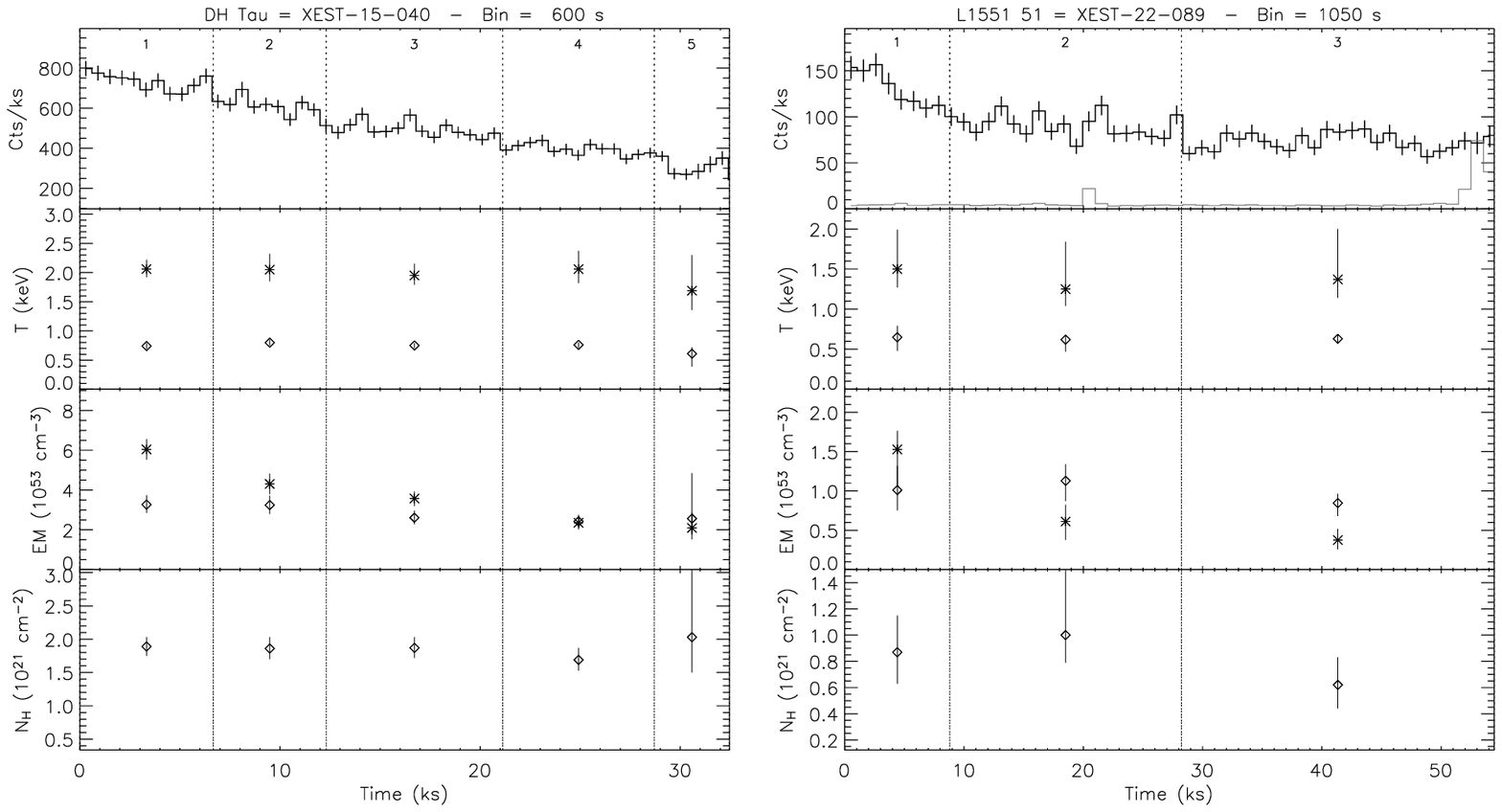}
\caption{Same as Fig.~\ref{longdec1} for DH~Tau (XEST-15-040) and L1551~51
(XEST-22-089).}
\label{longdec2}
\end{figure*}

\subsection{Very long decays or increases of the count rate}
\label{contdec}

A group of six sources in our sample showed a monotonic decrease or increase
of the count rate over the entire observation. One of them, XZ~Tau, has a
rising light curve, while for the other five sources (V830~Tau, HQ~Tau,
FS~Tau, DH~Tau and L1551~51) a long-lasting decline of the count rate was
observed. The light curves and best-fit parameters of this group of sources
are shown in Figs.~\ref{longdec1} and \ref{longdec2}.

\subsubsection{XZ~Tau (XEST-22-047)}

\object{XZ~Tau} (XEST-22-047) is a binary star composed of two CTTS of
spectral type M2 and M3.5. This source shows a long monotonic rise of the
count rate by a factor of $\sim$\,4 in $\sim$\,40~ks. An analysis of this
observation has been reported by \citet{Favata03} and \citet{Giardino06}. 

The quiescent emission at the beginning of the observation is rather hot,
with temperatures of 8 and 40~MK and $EM_2/EM_1\sim 0.7$. The cool component
does not vary significantly, except for an increase of the EM in the last
block. The hotter temperature peaks at $\sim 70$~MK in blocks \#2 and 3,
although the increase is not much significant due to the large errors, while
the EM continues to increase as the count rate increases up to the last
block. Our results are in agreement with those reported by
\citet{Giardino06}. We note that the trend of $T_2$ and $EM_2$ resembles
that observed during the first part of the rising phase of V410~Tau,
suggesting that we are observing a similar event. 

\subsubsection{V830~Tau (XEST-04-016)}

\object{V830~Tau} (XEST-04-016, WTTS, spectral type K7) shows a long-lasting
decay by a factor of 3 in $\sim$\,30~ks. The two temperatures remain nearly
constant, within the errors, around $\sim$\,7 and 20~MK during the entire
decay (apart from a possible increase in block \#3), and the observed
decrease of the count rate is due only to the decrease of the two EMs with
time. We note that also the average plasma temperature (weighted with the
EMs) does not vary during the decay. It is possible that we are observing
the final stage of a long-lasting flare, when the temperature has already
returned to the pre-flare level while the EM is still changing. An
alternative explanation might be rotational modulation: the relative
amplitude of the count rate variation around the average count rate is $\sim
60$\%, compatible with the range of 20\,--\,70\% found in the Orion Nebula
Cluster \citep{Flaccomio05}. However, given the high amplitude of the
variation, and the fact that the 30~ks exposure covers only one tenth of the
star's rotational period \citep[$P_\mathrm{rot}=2.75$~d, see Table~10
in][]{Guedel06}, we believe that this interpretation is unlikely, although
we cannot draw definitive conclusions from the available data.

\subsubsection{HQ~Tau (XEST-09-026)}

\object{HQ~Tau} (XEST-09-026) is a close binary WTTS. Its light curve shows
a long-lasting decay, with a decrease of the count rate by a factor of 2 in
30~ks. The spectral analysis shows a significant decrease of the hotter
temperature, from 34 to 20~MK, and of both EMs, while the temperature of the
cool component remains constant. This behaviour suggests that we are
observing the end of the decay of a long-duration flare. The e-folding
timescale of the observed decay is $\sim$\,45~ks. From the decay of the
temperature and EM of the hottest component we find a high slope
$\zeta\sim$\,2, compatible with a freely-decaying loop. Using
Eq.~(\ref{l_free}) we obtain $L\ga 7\times 10^{11}$~cm. Since the stellar
radius of this star is not known, we cannot determine whether this
represents a compact or a large loop.

\subsubsection{FS~Tau (XEST-11-057)}

\object{FS~Tau} (XEST-11-057) is a binary composed of two CTTS of spectral
type M0 and M3.5. Its light curve decreases by a factor of $\sim 6$ during
the first 20~ks, with an e-folding timescale of $15 \pm 2$~ks, and then
remains at a low quiescent level for the remaining 20~ks. In this case the
spectra are well described by just one temperature, the low-temperature
component being unconstrained due to the high absorption ($N_\mathrm{H}\sim
1 - 1.6\,\times 10^{22}$~cm$^{-2}$). The temperature does not change
significantly, decreasing from 40~MK to 35~MK from the beginning to the end
of the observation. On the other hand the EM decreases significantly; a
small but significant decrease of the absorption is also observed. In this
case, it is not possible to apply the \citet{Reale97} method, since the rate
of decrease of the temperature is too low ($\zeta=0.2$). Applying the
formula for a freely-decaying loop we estimate an upper limit to the loop
length of $L\sim 2.7\times 10^{11}$~cm ($\sim\, 4\,R_\star$). Note that,
although the peak temperature of the flare is likely higher and therefore
the above upper limit is underestimated, we expect the true loop size to be
much smaller because of the presence of strong sustained heating, as
indicated by the shallow slope $\zeta$. 

\subsubsection{DH~Tau (XEST-15-040)}

\object{DH~Tau} (XEST-15-040) is a binary CTTS with a separation of
2.3$\arcsec$, unresolved by XMM-{\it Newton}. It shows a monotonic decay by
a factor of $\sim 3$ over the 30~ks exposure. As for V830~Tau, the two
temperatures are steady during the decay, at $\sim$\,9 and $\sim$\,25~MK,
while the two EMs decrease. In this case, contrary to V830~Tau, also the
ratio $EM_2/EM_1$ decreases, implying that, as time proceeds, less and less
hot material contributes to the emission. This suggests that we are
observing the end of a long-lasting flaring event. As for the previous
source, very strong sustained heating is present ($\zeta=0.2$), preventing
the use of the \citet{Reale97} method, and we estimate $L<4.5\times
10^{11}$~cm $\sim 3.5\,R_\star$.

\subsubsection{L1551~51 (XEST-22-089)}

\object{L1551~51} (XEST-22-089, WTTS) also shows a decay by a factor of 2
during the first 30~ks, then stays at a constant level for the remaining
20~ks of observation. Also for this source the temperatures do not vary, and
the observed variation is due only to the changing EMs. As for DH~Tau, also
the EM ratio decreases, implying a decrease of the average plasma
temperature, likely due to a flare decay. This source also shows very strong
sustained heating ($\zeta=0.16$). From the decay e-folding time of 9.5~ks we
estimate $L\la 1\times 10^{11}$~cm, comparable to the stellar radius.

\begin{figure*}
\centering
\resizebox{\hsize}{!}{\includegraphics[width=\hsize]{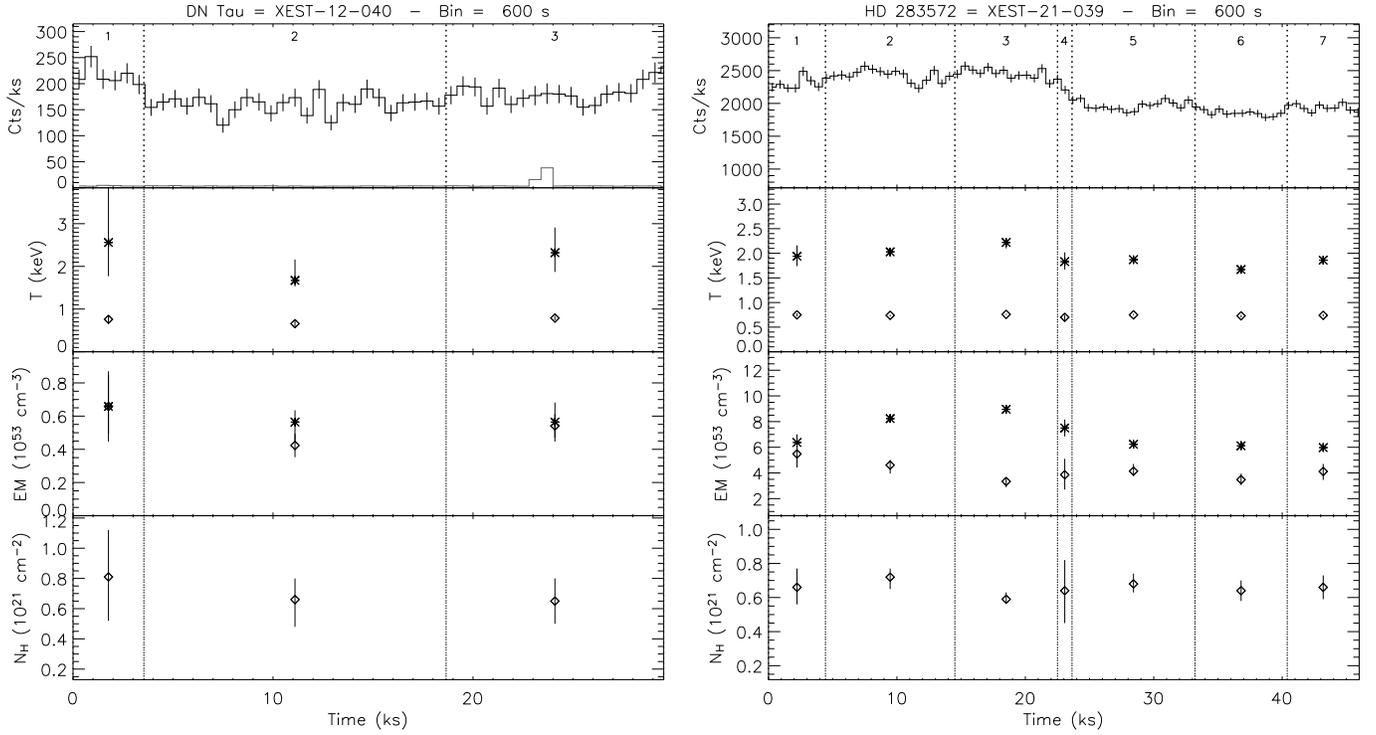}}
\caption{Light curve and spectral fitting results for the TMC sources DN~Tau
(XEST-12-040) and HD~283572 (XEST-21-039), showing a slow modulation. Panels
are the same as in Fig.~\ref{impfl}.}
\label{slowvar} 
\end{figure*}

\subsection{Other types of variability}
\label{othervar}

\begin{figure}
\centering
\resizebox{\hsize}{!}{\includegraphics{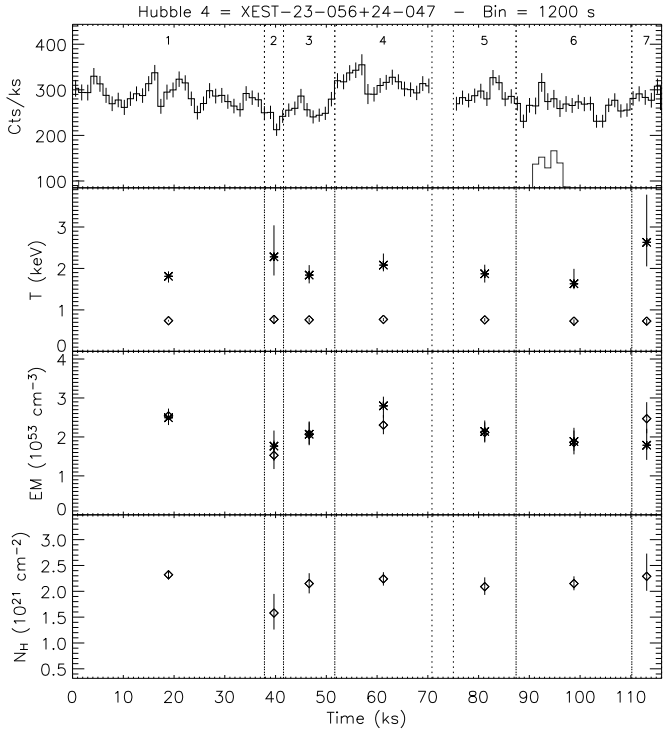}}
\caption{Light curve and spectral fitting results for the TMC source
Hubble~4 (XEST-23-056/24-047) showing a complex variability. Panels are the
same as in Fig.~\ref{impfl}.}
\label{complex}
\end{figure}

The remaining three XEST sources identified with TMC members show
significant variability in the form of slow modulation (DN~Tau and
HD~283572, Fig.~\ref{slowvar}) or complex flare-like variability (Hubble~4,
Fig.~\ref{complex}).

\subsubsection{DN~Tau (XEST-12-040)}

\object{DN~Tau} (XEST-12-040, CTTS, spectral type M0) shows a modulated
light curve with a higher count rate at the beginning and the end of the
exposure and a minimum in the second time block; the amplitude of the
modulation with respect to the mean level is $\sim$\,15\%. The plasma
parameters do not show significant variations, being consistent within the
errors, although there is a tendency for higher $T_2$ and $EM_1$ in the
first and last block.

\subsubsection{HD~283572 (XEST-21-039)}

\object{HD~283572} (XEST-21-039) is a G5 WTTS with $P_\mathrm{rot} =
1.55$~d; the present observation, with an exposure time of $\sim 45$~ks,
therefore covers about one third of the rotational period. The light curve
shows a clear modulation, with two characteristic levels of emission. During
the first 20~ks the count rate increases slightly, then in block \#4 it
decreases reaching a level a factor of $\sim 1.3$ lower, that is maintained
until the end of the observation. The two temperatures do not vary
significantly, with the cooler one staying around 8\,--\,9~MK, and the
hotter one varying between 20 and 25~MK with a trend that closely follows
that of the light curve. On the other hand, the two EMs show significant
variations, in particular in the first part of the observation, where the
cool EM decreases and the hot one increases, leading to an increase in their
ratio and therefore in the relative contribution of the hotter plasma to the
emission. These characteristics suggest that the observed variability is
likely due to the appearing and disappearing, due to the stellar rotation,
of active regions containing hotter plasma with respect to the rest of the
corona. 

\subsubsection{Hubble~4 (XEST-23-056/24-047)}

The light curve of \object{Hubble~4} (XEST-23-056/24-047, WTTS, spectral
type K7) shows significant low-level flare-like variability superimposed on
a slow modulation. The average count rate shows a very slow decrease during
the first 50~ks, then it increases by a factor of $\sim$\,1.5 at the
beginning of block \#4, and decays again slowly until the end of the
observation. The two temperatures do not vary significantly, having average
values of $\sim$\,9 MK and $\sim$\,20~MK, with the exception of the last
block where the hot component is slightly higher. As in other sources, the
variations of the two EMs are more significant, showing a modulation with
time of both values, but keeping their ratio nearly constant, between 1 and
1.2. The observed slow variation of the mean count rate might be due to
rotational modulation of the emission from active regions rotating in and
out of view.

\section{Sources not associated with known TMC members}
\label{nonmembers}

\begin{figure*}
\centering
\includegraphics[width=\hsize]{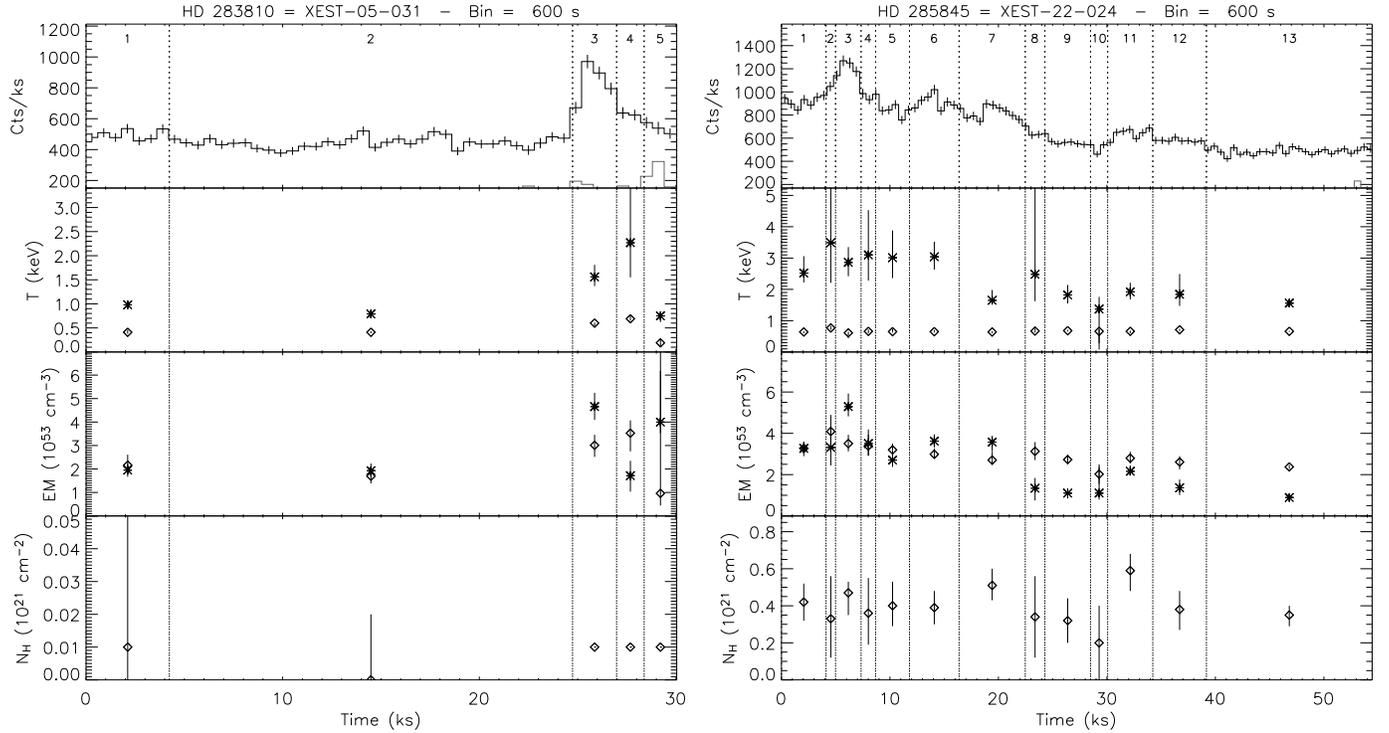}
\caption{Light curve and spectral fitting results for the XEST sources
HD~283810 (XEST-05-031) and HD~285845 (XEST-22-024), not associated with the
TMC. Panels are the same as in Fig.~\ref{impfl}.}
\label{nonm}
\end{figure*}

In addition to the TMC members, we have studied three additional XEST
sources (XEST-05-031 = HD~283810, XEST-16-031 = 2MASS~J04195676+2714488, and
XEST-22-024 = HD~285845) that are likely non-members of the TMC but that are
strong X-ray sources and show significant variability. Their light curves
and spectral parameters are shown in Figs.~\ref{nonm} and \ref{xest16031}.

\subsection{HD~283810 (XEST-05-031)}

XEST-05-031 is identified with \object{HD~283810}, a K5V star with H$\alpha$
emission, which is probably an older foreground star since it has radial
velocity inconsistent with TMC membership and does not show significant Li
absorption \citep{Herbig86}; assuming a main-sequence object, its photometry
($V=10.74$, $B-V=1.03$) locates it at a distance of $\sim$\,60~pc. Its light
curve shows a flare at the end of the observation, with an increase of the
count rate by a factor of 2.5 and an e-folding decay time of $\sim$\,2.4~ks.
The hotter temperature rises from $\sim$\,9~MK during quiescence to 18~MK in
block \#3 and reaches a maximum value of 26~MK in block \#4, after the flare
peak. Both EMs increase significantly at the flare peak. Given the irregular
trend of the temperature and EM, we cannot apply the \citet{Reale97} method;
using the formula for a freely-decaying loop and the observed temperature in
block \#3, we derive a loop semi-length of $L<3 \times 10^{10}$~cm, which is
comparable to the stellar radius for a K5 main-sequence star
\citep{Siess00}.

\subsection{HD~285845 (XEST-22-024)}

XEST-22-024 is identified with \object{HD~285845}, which is a foreground
binary system at a distance of 90~pc; the absence of the Li absorption line
at 6708~\AA\ indicates that the system is not composed of PMS stars
\citep{Walter88,Favata03}. The mean properties of the X-ray spectrum have
been discussed by \citet{Favata03}, who reported a spectral fitting with a
2-T model characterized by high Ca and Ne abundances with respect to a Fe
abundance around $0.26\,Z_\odot$.  

The light curve of this source shows a complex variability. A flare occurred
at the beginning of the exposure, with a gradual rise lasting $\sim$\,5~ks,
followed by a decay on which several minor impulses are superimposed. The
initial decay just after the peak has an e-folding time of $\sim\,2$~ks. The
emission reaches a steady level $\sim 30$~ks after the peak. As observed in
other sources, the cool temperature does not vary significantly during the
flare evolution. The hotter temperature peaks in block \#2, during the rise
phase, at 40~MK, and remains steady at a level of $\sim$\,30\,--\,35~MK up
to block \#7, when it decreases to $\sim$\,15\,--\,25~MK, remaining at this
level until the end of the observation. The observed light curve variations
are mostly due to the EMs of both components, that vary significantly
throughout the observation. Given the complexity of the light curve, it is
not possible to derive the loop size by fitting the $T$ vs $EM$ decay. We
can make a rough estimate assuming the initial decay time of 2~ks and
Eq.~(\ref{l_free}), obtaining $L\sim 3\times 10^{10}$~cm. Assuming a G6
main-sequence star, this corresponds to $\sim 0.4\,R_\star$ \citep{Siess00}.

\begin{figure*}
\centering
\includegraphics[width=\hsize]{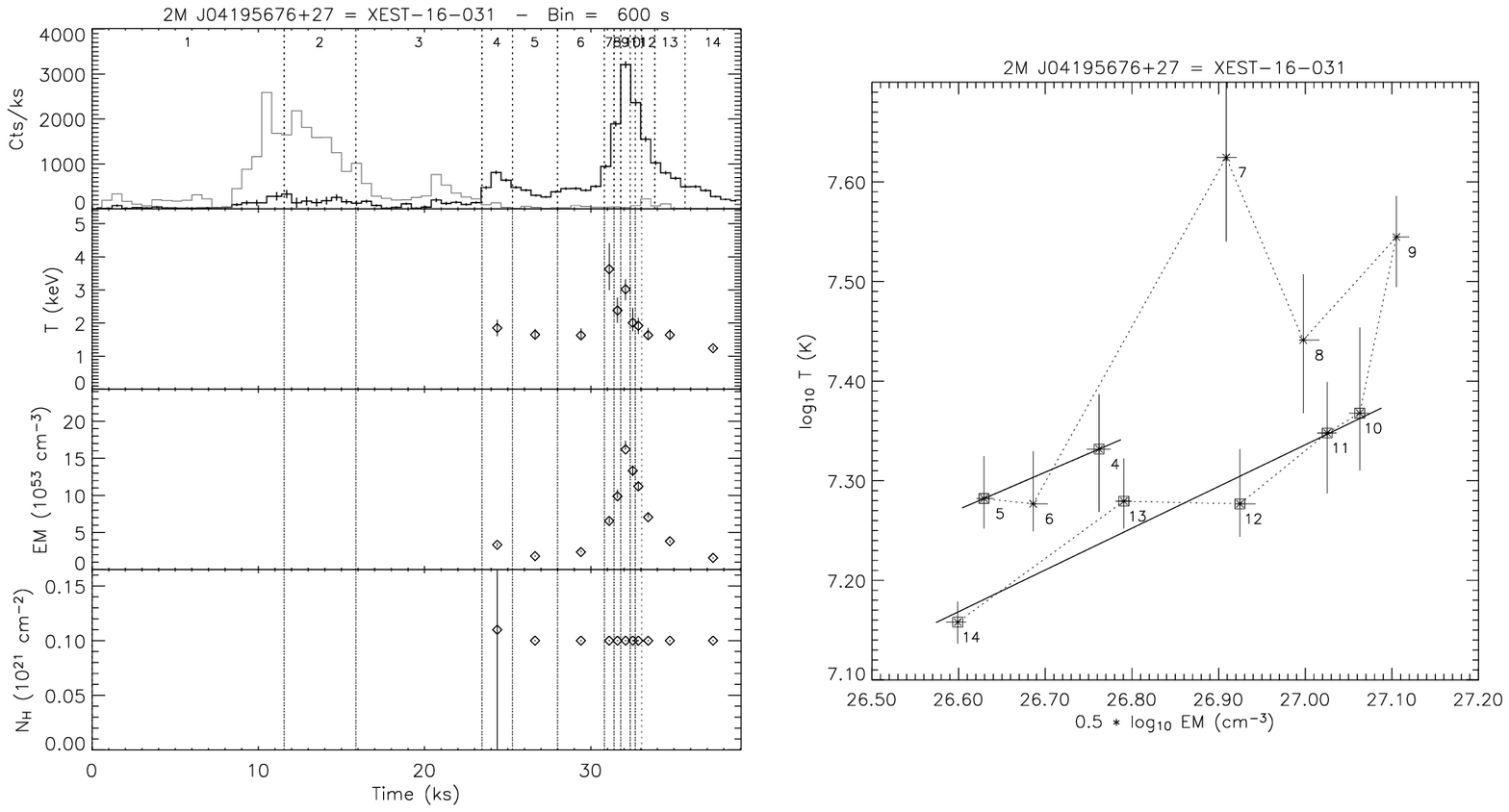}
\caption{{\em Left}: Light curve and spectral fitting results for
XEST-16-031. Panels are the same as in Fig.~\ref{impfl}. {\em Right}:
Evolution of the best-fit temperature and EM of the hot component. See
Fig.~\ref{xest26072} for details. The slope has been fitted between points
\#10 and \#14. The slope during the decay of the first small flare (points
\#4 and \#5) is also indicated.}
\label{xest16031}
\end{figure*}

\subsection{2MASS~J04195676+2714488 (XEST-16-031)}

Source XEST-16-031 has an IR counterpart in the 2MASS catalogue,
\object{2MASS~J04195676+2714488}, located at $\sim 0.2\arcsec$ and with
magnitudes $J=12.38$, $H=11.80$ and $K=11.54$ mag; another fainter 2MASS
source is present at 4.6$\arcsec$, however the X-ray source is most likely
associated with the former one, given the accuracy of the XEST positions
\citep[see][]{Guedel06,Scelsi06}. 2MASS~J04195676+2714488 has photometry
inconsistent with TMC membership, being located on the main sequence in the
colour-magnitude diagrams for the distance of the TMC \citep{Scelsi06}, and
is therefore likely to be a field late-type star. The uncertainties in the
photometry allow a distance between 80 and 190~pc for a main sequence star.

During the first 23~ks, the source was in a quiescent state with very low
emission. Unfortunately in this part of the observation (blocks \#1 to 3)
the background was very high and dominating the observed count rate (see
Fig.~\ref{xest16031}), preventing the possibility of performing spectral
analysis. In block \#4, $\sim$\,24~ks after the start of the observation, a
flare occurred, followed by a second much stronger flare 7~ks later. The
first flare has a peak count rate a factor of $\sim$\,5 higher than the
quiescent level, while the second flare increases the count rate by more
than one order of magnitude. The two flares have very similar decay times,
with e-folding times of $\sim$\,1.6 and 1.5~ks, respectively. This suggests
that they may have occurred in loops of comparable size.

The absorption is very low, consistent with a column density $N_\mathrm{H} =
1 \times 10^{20}$~cm$^{-2}$, and to better constrain the fit parameters it
was kept fixed to this value. The spectrum was well fitted with only one
temperature component. Since the emission during the quiescent phase is
negligible with respect to the flare phase, we assume that the 1-T model is
representing essentially the flaring plasma during the time blocks from \#4
to \#14.

The right panel of Fig.~\ref{xest16031} shows the evolution of the
temperature and the EM during the observation. In block \#4, corresponding
to the peak of the first, small flare, the plasma temperature is 22~MK; the
temperature and EM decrease in block \#5 during the flare decay, and remains
steady in the following time interval. In block \#7, at the start of the
rise of the second flare, the temperature increases to $\sim$\,42~MK; a
second re-heating is observed in block \#9, corresponding to the flare peak,
where the EM reaches its maximum value. The flare cooling phase from point
\#10 to point \#14 proceeds with a slope $\zeta = 0.42\pm 0.07$, that
indicates the presence of significant sustained heating after the initial
ignition. It is worth noting that the slope between points \#4 and \#5,
relative to the decay of the first small flare, has a very similar value of
0.37. Using Eq.~(\ref{l_loop}) we derive $L\sim 1\times 10^9$~cm for the
first flare, and $L \sim 5\pm 4 \times 10^9$~cm for the second flare. These
values are significantly smaller than those estimated for the other sources.
Although the precise nature of this source cannot be assessed here, given
the very low absorption we suggest that it could be an older M-type star
located just in front of the cloud and showing flare events analogous to
those observed on the Sun and active late-type stars.

\setcounter{table}{2}
\begin{table*}
\centering
\caption{Summary of the parameters derived for the XEST sources. For all 
sources we give the temperature of the hot component ($T_\mathrm{q}$) and
the luminosity ($L_\mathrm{Xq}$) in the quiescent (or lowest count rate)
intervals; a range is given for sources with slow modulation or complex
variability. For sources showing flares or prolonged rise/decays, we also
give the peak temperature ($T_\mathrm{peak}$) and luminosity
($L_\mathrm{Xpeak}$), the rise and/or decay e-folding timescales
($\tau_\mathrm{rise}$ and $\tau_\mathrm{dec}$), and the derived loop size
$L$.}
\label{tab_summ}
\begin{tabular}{llccrccrcc}
\hline\hline\noalign{\smallskip}
XEST ID& Optical ID & $T_\mathrm{q}$& $L_\mathrm{Xq}$&
$T_\mathrm{peak}$& $L_\mathrm{Xpeak}$& $\tau_\mathrm{rise}$&
$\tau_\mathrm{dec}$& $L$& $L/R_\star$\\[2pt]
   &  & (MK)& ($10^{30}$\,erg\,s$^{-1}$)& (MK)& ($10^{30}$\,erg\,s$^{-1}$)& (ks)& (ks)& 
($10^{11}$\,cm)& \\
\noalign{\smallskip}\hline\noalign{\smallskip}
04-016       & V830 Tau    & 22      & 2.8     & $\ge$\,33& $\ge$\,8.5& $\cdots$& 11.4    & $\cdots$   & $\cdots$\\
09-026       & HQ Tau      & 19      & 4.6     & $\ge$\,34& $\ge$\,9.1& $\cdots$& 45.2    & $\ga$\,7.1 & $\cdots$\\
11-057       & FS Tau      & 36      & 1.4     & $\ge$\,42& $\ge$\,6.3& $\cdots$& 15.2    & $\la$\,2.7 & 4.1     \\
12-040       & DN Tau      & 19--27  & 1.0--1.4& $\cdots$ & $\cdots$  & $\cdots$& $\cdots$& $\cdots$   & $\cdots$\\
15-040       & DH Tau      & 20      & 4.7     & $\ge$\,24& $\ge$\,9.8& $\cdots$& 34.2    & $\la$\,4.5 & 3.5     \\
17-066       & JH 108      & 28      & 1.5     & \ \,46   & \ \ 3.7   & \ 3.8   & \ 3.3   & $\la$\,0.6 & 0.7     \\
21-039       & HD 283572   & 19--26  & 10--14  & $\cdots$ & $\cdots$  & $\cdots$& $\cdots$& $\cdots$   & $\cdots$\\
22-047       & XZ Tau      & 40      & 0.9     & \ \,70   & \ \ 3.8   & 35.3    & $\cdots$& $\cdots$   & $\cdots$\\
22-089       & L1551 51    & 16      & 1.4     & $\ge$\,17& $\ge$\,2.5& $\cdots$& \ 9.5   & $\la$\,1.1 & 1.1     \\
23-032/24-028& V410 Tau    & 20      & 3.8     & \ \,43   &  11.7     & 20.9    & \ 8.1   & 0.1--0.7   & 0.1--0.4\\
23-047/24-040& V892 Tau    & 23      & 9.5     & \,100    &  46.8     & \ 3.4   & $\cdots$& 1.0$^{\,a}$& 0.5    \\
23-050/24-042& V410 X7     & 24      & 0.7     & $\ge$\,63& $\ge$\,4.9& \ 1.5   & \ 2.1   & $\ge$\,0.5 & 0.4     \\
23-056/24-047& Hubble 4    & 19--30  & 3.5--5.4& $\cdots$ & $\cdots$  & $\cdots$& $\cdots$& $\cdots$   & $\cdots$\\
23-074/24-061& V819 Tau    & 14      & 2.0     & \ \,19   & \ \ 3.8   & \ 1.3   & 10.1    & 0.7        & 0.5     \\
26-072       & HBC 427     & 25      & 3.2     & \ \,72   &  15.8     & \ 3.2   & 20.2    & 2.4        & 1.9     \\
28-100       & BP Tau      & 18      & 1.5     & \ \,47   &  \ \,4.9  & \ 2.6   & 10.4    & 1.2        & 0.9     \\
\noalign{\smallskip}\hline\noalign{\smallskip}
05-031       & HD 283810   & 10      & 1.5$^{\,b}$& \ \,26& \ \ 3.2$^{\,b}$& \ 0.5& \ 2.4 & 0.3        & $\cdots$\\
16-031       & J04195676+27& $\cdots$& $\cdots$& \ \,21   & \ \ 3.5   & \ 0.6   & \ 1.6   & 0.01       & $\cdots$\\
\ \ \ ''     & \ \ \ \ ''  & $\cdots$& $\cdots$& \ \,42   &  20.0     & \ 0.8   & \ 1.5   & 0.01--0.09 & $\cdots$\\
22-024       & HD 285845   & 18      & 2.1$^{\,c}$& \ \,40& \ \ 6.5$^{\,c}$& $\cdots$& $\cdots$& $\cdots$& $\cdots$\\
\noalign{\smallskip}\hline\noalign{\smallskip}
\multicolumn{10}{l}{$^{\,a}$: from \citet{Giardino04}}\\
\multicolumn{10}{l}{$^{\,b}$: computed for $d = 60$ pc}\\
\multicolumn{10}{l}{$^{\,c}$: computed for $d = 90$ pc}\\
\end{tabular}
\end{table*}

\section{Discussion and conclusions}
\label{concl}

In this paper we have studied a sample of 19 bright variable X-ray sources
detected in the XEST survey. Our sample includes 16 known TMC members (10
WTTS, 5 CTTS and a Herbig~Ae star) plus three additional sources unrelated
to the cloud but showing flaring events. The studied sources show different
types of variability, in the form of flares, either with fast rise and
slower decay or with symmetrical shapes, continuous rise or decays, slow
modulation possibly due to rotation, and complex variability, with
flare-like events superimposed on a slow modulation or decay. Using detailed
time-resolved spectroscopy we have investigated the changes of the
parameters of the emitting plasma (temperatures and emission measures) and
of the hydrogen column density, and, in the case of flares, we have derived
informations on the size of the involved coronal structures. The main
parameters derived for the studied sources are summarized in
Table~\ref{tab_summ}. 

The quiescent emission has typical temperatures $T_1\sim 4-10$~MK and
$T_2\sim 15-35$~MK, with $EM_2/EM_1 \sim 0.7-2$, consistent with the values
found in other studies of young PMS stars \citep[e.g][]{Feigel99,Wolk05}. No
significant difference is evident in our small sample between the spectral
characteristics of CTTS and WTTS: both the ranges of the parameters and
their median values are similar for the two classes. The Herbig~Ae star
V892~Tau has spectral characteristics very similar to those found for TTS
stars, in agreement with other studies \citep{Hamaguchi05,Stelzer06hae},
supporting the suggestion that the emission might come from its cool,
unresolved close companion rather than from the Herbig~Ae star itself. We
find that in most sources the cool plasma component does not vary
significantly, even during flares, while the observed time evolution can be
totally ascribed to variations in the hot component. Similar results have
been obtained for the Orion PMS stars studied in the COUP survey
\citep{Wolk05}, as well as for older active stars \citep[e.g.][]{Audard01}.

Nine of the studied sources (including two TMC non-members) show evident
flaring activity. The flares observed on TMC members have total duration
between $\sim$\,20 and 70~ks ($\sim 5-20$~hr), with e-folding rise
timescales of $\sim 1-4$~ks, except for the peculiar flare on V410~Tau with
$\tau_\mathrm{rise}\sim\,20$~ks, and decay timescales of $\sim 2-20$~ks. An
additional 5 sources show gradual decays over 30\,--\,50~ks, that might
represent the decay of long-lasting flares, as suggested by the decreasing
plasma temperature and emission measure, and one source shows a prolonged
rise (with e-folding timescale of $\sim$\,35~ks) with spectral
characteristics similar to those observed during the long rise phase of the
atypical flare on V410~Tau. Unfortunately, the typical 30\,--\,40~ks
exposure times of the XEST survey introduce a bias against the detection of
long-duration flares in our observations. Indeed, we detected flares with
total duration of $\sim\,50\--\,70$~ks in the archival fields having
exposure times of $\sim$\,100~ks. It is conceivable that the sources with
gradual decays were undergoing flares of similar duration or even longer,
such as found in the COUP survey, where events lasting up to 3 days were
observed \citep{Favata05,Wolk05}. 

Spectral analysis of the flaring sources shows peak temperatures from 40~MK
up to 100~MK for the strongest flares, and peak luminosities between $4
\times 10^{30}$ and $5 \times 10^{31}$~erg~s$^{-1}$. On the other hand, the
flares observed on the two TMC non-members show significantly shorter rise
and decay times ($\la 1$~ks and $\sim$\,2~ks, respectively), and peak
temperature in the range 20\,--\,40~MK.

For four of the flaring sources (the WTTS HBC~427 and V410~Tau, the CTTS
BP~Tau, and the non-member 2MASS~J04195676+2714488) we could perform a
detailed analysis of the decay phase using the method by \citet{Reale97};
the method was applied also to V819~Tau, for which however only two points
in the decay are available. In all these cases we find that significant
residual heating must be present during the decay, governing the observed
light curve evolution. For the other flaring sources, we do not have enough
intervals in the decay phase to apply the method, and we could only estimate
upper limits to the loop size using the formula for a freely-decaying loop.
The loop size is fully constrained only for the CTTS BP~Tau, which has $L =
1.2 \times 10^{11}$~cm, comparable to the stellar radius, and for the WTTS
HBC~427, which has $L = 2.4\times 10^{11}$~cm corresponding to
$\sim\,2\,R_\star$; the latter star has the longest decay time, and hence
the longest loop size, among the sources for which the entire flare
evolution is observed. The other TMC members have loop lengths in the range
$4-7\times 10^{10}$~cm, smaller than or comparable to the stellar radius.
For the sources with continuous decay, only one (HQ~Tau) has a steep slope,
compatible with a freely-decaying loop with no additional heating after the
initial ignition; for this star we find $L\sim 7\times 10^{11}$~cm. The
other four sources have upper limits between $1 - 5\times 10^{11}$~cm,
comparable to the stellar radius for the two WTTS, but equal to $\sim
4\,R_\star$ for the two CTTS (FS~Tau and DH~Tau). Finally, for the TMC
non-members we find loop sizes in the range $1\times 10^9 - 3 \times
10^{10}$~cm.

We mention that flare characteristics similar to those found here for TMC
members are observed also for the CTTS SU~Aur
\citep[XEST-26-067,][]{Francio07su}, that showed three flares during the
observation with rise and decay times of $\sim$\,6~ks and
$\sim$\,5\,--\,9~ks, respectively, and peak temperatures of
$\sim$\,50\,--\,140~MK. The data do not allow a detailed flare analysis,
however using the parameters reported by \citet{Francio07su} we derive upper
limits to the loop semi-length of $\sim\,1.1-1.8 \times 10^{11}$~cm,
comparable to the stellar radius.

Previous observations of PMS stars have shown the presence of compact
flaring structures, with $L\la R_\star$, similar to what is observed in
active late-type stars
\citep{Favata01,Grosso04,Giardino06,Guedel04prox,Reale04}. \citet{Favata05}
analyzed a sample of strong flares detected in the COUP survey, finding both
compact structures, of size shorter than a stellar radius, and very extended
structures, with lenghts of $5 - 20\,R_\star$. Such long structures possibly
represent magnetic loops connecting the stellar surface with the
circumstellar disk. Our sample shows generally loops of size comparable to
or smaller than the stellar radius: the longest loop with a fully
constrained size is about 2 stellar radii in length, which is compatible
with a loop anchored on the stellar surface. We stress that this loop has
been observed on a WTTS, that should not possess a circumstellar disk, and
that a similar loop size ($\sim\,1.6\,R_\star$) has been found also on the
WTTS V827~Tau by \citet{Giardino06}. A possible hint for large loops of size
$\sim 4\,R_\star$ is found for the two CTTS with long-lasting decays,
suggesting that these stars might have indeed loops connecting the star and
the disk; however the estimated size is highly uncertain, since we are
observing only a small part of the decay, therefore we cannot draw any
definitive conclusion on the size of the emitting structures.

Finally, we have studied two sources showing possible rotational modulation
with amplitudes of $\sim 15$\%, and two other sources (one unrelated to the
TMC) with flare-like variability superimposed on a slow modulation or on a
flare decay. Except for the last case, we do not find significant variations
in the plasma temperatures, and the observed variability is mainly
determined by variations of the emission measures.

\begin{acknowledgements}
EF, IP and GM would like to thank F. Reale for useful discussions on the
interpretation of flare variability. We acknowledge financial support by the
International Space Science Institute (ISSI) in Bern, Switzerland to the
XMM-{\it Newton} XEST team. The Palermo group acknowledges financial
contribution from contract ASI-INAF I/023/05/0. X-ray astronomy research at
PSI has been supported by the Swiss National Science Foundation (grants
20-66875.01 and 20-109255/1).
This research is based on observations obtained with XMM-{\it Newton}, an
ESA science mission with instruments and contributions directly funded by
ESA Member States and NASA. 
\end{acknowledgements}

\bibliographystyle{aa}
\bibliography{bib_paper}

\longtab{2}{
\tiny
\begin{longtable}{llrrrrrrrrcc}
\caption{\label{fit} Best fit values of the spectral model parameters.
Errors are 90\% confidence ranges for one interesting parameter; where
errors are not given, the parameter was held fixed to the tabulated value.
$F_\mathrm{X}$ and $L_\mathrm{X}$ are the unabsorbed X-ray flux and
luminosity in the 0.3--7.3 keV band.
$L_\mathrm{X}$ is computed from $F_\mathrm{X}$ assuming a distance of
140 pc for all stars, except for HD~283810 and HD~285845, where we used the
photometric distance of 60 and 90 pc, respectively.
Note that for J04195676+27 (XEST-16-031) the reported luminosity must be
considered only indicative, since the true distance of this star is not
known.
}\\
\hline\hline \noalign{\smallskip}
XEST-ID & Optical ID& Block& \multicolumn{1}{c}{$N_\mathrm{H}$}&
   \multicolumn{1}{c}{$T_1$}& \multicolumn{1}{c}{$T_2$}&
   \multicolumn{1}{c}{$EM_1$}& \multicolumn{1}{c}{$EM_2$}&
   $\chi_\mathrm{r}^2$& dof& \multicolumn{1}{c}{$F_\mathrm{X}/10^{-12}$} &
   \multicolumn{1}{c}{$\log L_\mathrm{X}$} \\[2pt]
& & \# & $10^{21}$\,cm$^{-2}$& \multicolumn{1}{c}{keV}&
   \multicolumn{1}{c}{keV}& \multicolumn{2}{c}{$10^{53}$\,cm$^{-3}$}& 
   & & \multicolumn{1}{c}{erg\,cm$^{-2}$\,s$^{-1}$} &
   \multicolumn{1}{c}{erg\,s$^{-1}$} \\
\noalign{\smallskip}\hline\noalign{\medskip}
\endfirsthead
\caption{continued}\\
\hline\hline\noalign{\medskip}
XEST-ID & Optical ID& Block& \multicolumn{1}{c}{$N_\mathrm{H}$}&
  \multicolumn{1}{c}{$T_1$}& \multicolumn{1}{c}{$T_2$}&
  \multicolumn{1}{c}{$EM_1$}& \multicolumn{1}{c}{$EM_2$}&
  $\chi_\mathrm{r}^2$& dof& \multicolumn{1}{c}{$F_\mathrm{X}/10^{-12}$} &
  \multicolumn{1}{c}{$\log L_\mathrm{X}$} \\[2pt]
& & \# & $10^{21}$\,cm$^{-2}$& \multicolumn{1}{c}{keV}&
   \multicolumn{1}{c}{keV}& \multicolumn{2}{c}{$10^{53}$\,cm$^{-3}$}&
   & & \multicolumn{1}{c}{erg\,cm$^{-2}$\,s$^{-1}$} &
   \multicolumn{1}{c}{erg\,s$^{-1}$} \\
\noalign{\smallskip}\hline\noalign{\medskip}
\endhead
\noalign{\medskip}\hline
\endfoot
\noalign{\medskip}\hline
\endlastfoot
04-016& V830 Tau& 
      1& $0.33_{-0.11}^{+0.10}$  & $0.79_{-0.07}^{+0.06}$    & $2.08_{-0.25}^{+0.44}$& $2.30_{-0.49}^{+0.45}$    & $5.64_{-0.66}^{+0.66}$ & 1.06& 289& \ 3.59& 30.93\\[2pt]
 & &  2& $0.34_{-0.24}^{+0.25}$  & $0.51_{-0.11}^{+0.24}$    & $1.88_{-0.25}^{+0.26}$& $1.29_{-0.42}^{+0.54}$    & $5.08_{-0.73}^{+0.54}$ & 1.21&  99& \ 2.80& 30.82\\[2pt]
 & &  3& $0.46_{-0.16}^{+0.26}$  & $0.98_{-0.21}^{+0.08}$    & $2.80_{-1.19}^{+2.66}$& $2.85_{-1.51}^{+0.82}$    & $2.63_{-0.92}^{+2.38}$ & 1.09&  96& \ 2.56& 30.78\\[2pt]
 & &  4& $0.33_{-0.12}^{+0.12}$  & $0.67_{-0.06}^{+0.07}$    & $1.86_{-0.20}^{+0.25}$& $1.41_{-0.24}^{+0.21}$    & $2.87_{-0.35}^{+0.28}$ & 1.12& 193& \ 1.87& 30.64\\[2pt]
 & &  5& $0.32_{-0.13}^{+0.14}$  & $0.76_{-0.09}^{+0.06}$    & $1.95_{-0.31}^{+0.49}$& $1.25_{-0.26}^{+0.28}$    & $2.07_{-0.35}^{+0.19}$ & 1.16& 143& \ 1.47& 30.54\\[2pt]
 & &  6& $0.24_{-0.18}^{+0.20}$  & $0.65_{-0.27}^{+0.11}$    & $1.93_{-0.30}^{+0.52}$& $0.96_{-0.21}^{+0.24}$    & $1.81_{-0.35}^{+0.28}$ & 1.10&  94& \ 1.21& 30.45\\
\noalign{\medskip}\hline\noalign{\medskip}
09-026& HQ Tau& 
      1& $4.94_{-1.12}^{+1.36}$  & $0.61_{-0.24}^{+0.16}$    & $2.90_{-0.62}^{+1.11}$& $4.89_{-1.98}^{+3.90}$    & $3.48_{-0.75}^{+1.22}$ & 0.80&  92& \ 3.88& 30.96\\[2pt]
 & &  2& $4.52_{-0.43}^{+0.49}$  & $0.76_{-0.06}^{+0.07}$    & $2.00_{-0.23}^{+0.36}$& $3.17_{-0.66}^{+0.85}$    & $3.25_{-0.54}^{+0.49}$ & 1.05& 238& \ 2.84& 30.82\\[2pt]
 & &  3& $3.58_{-0.60}^{+1.03}$  & $0.71_{-0.15}^{+0.12}$    & $1.92_{-0.35}^{+0.56}$& $1.98_{-0.66}^{+1.27}$    & $2.52_{-0.59}^{+0.66}$ & 0.98&  93& \ 1.97& 30.67\\[2pt]
 & &  4& $4.68_{-0.59}^{+0.74}$  & $0.74_{-0.14}^{+0.10}$    & $1.62_{-0.25}^{+0.71}$& $2.12_{-0.68}^{+0.73}$    & $2.42_{-0.85}^{+0.73}$ & 1.08&  83& \ 1.94& 30.66\\
\noalign{\medskip}\hline\noalign{\medskip}
11-057& FS Tau& 
      1& $15.99_{-1.90}^{+1.96}$ & $\ldots_{\phantom{-0.00}}$& $3.65_{-0.55}^{+0.78}$& $\ldots_{\phantom{-0.00}}$& $4.73_{-0.68}^{+0.73}$ & 0.91& 108& \ 2.68& 30.80\\[2pt]
 & &  2& $13.14_{-1.76}^{+1.87}$ & $\ldots_{\phantom{-0.00}}$& $3.61_{-0.60}^{+0.90}$& $\ldots_{\phantom{-0.00}}$& $2.87_{-0.42}^{+0.52}$ & 0.96&  82& \ 1.61& 30.58\\[2pt]
 & &  3& $11.18_{-1.32}^{+1.45}$ & $\ldots_{\phantom{-0.00}}$& $3.08_{-0.48}^{+0.58}$& $\ldots_{\phantom{-0.00}}$& $1.08_{-0.14}^{+0.19}$ & 0.93& 151& \ 0.59& 30.14\\
\noalign{\medskip}\hline\noalign{\medskip}
12-040& DN Tau& 
      1& $0.81_{-0.29}^{+0.31}$  & $0.76_{-0.10}^{+0.08}$    & $2.56_{-0.79}^{+5.11}$& $0.66_{-0.16}^{+0.19}$    & $0.66_{-0.21}^{+0.21}$ & 0.95&  65& \ 0.61& 30.16\\[2pt]
 & &  2& $0.66_{-0.18}^{+0.14}$  & $0.66_{-0.06}^{+0.10}$    & $1.67_{-0.14}^{+0.49}$& $0.42_{-0.07}^{+0.07}$    & $0.56_{-0.12}^{+0.07}$ & 1.15& 201& \ 0.42& 29.99\\[2pt]
 & &  3& $0.65_{-0.15}^{+0.15}$  & $0.79_{-0.05}^{+0.05}$    & $2.32_{-0.45}^{+0.59}$& $0.54_{-0.09}^{+0.07}$    & $0.56_{-0.09}^{+0.12}$ & 0.93& 169& \ 0.50& 30.07\\
\noalign{\medskip}\hline\noalign{\medskip}
15-040& DH Tau& 
      1& $1.89_{-0.14}^{+0.14}$  & $0.74_{-0.05}^{+0.05}$    & $2.06_{-0.14}^{+0.16}$& $3.27_{-0.42}^{+0.47}$    & $6.04_{-0.52}^{+0.52}$ & 0.96& 412& \ 4.18& 30.99\\[2pt]
 & &  2& $1.86_{-0.16}^{+0.17}$  & $0.80_{-0.05}^{+0.05}$    & $2.05_{-0.20}^{+0.27}$& $3.25_{-0.45}^{+0.47}$    & $4.30_{-0.52}^{+0.52}$ & 1.07& 298& \ 3.36& 30.90\\[2pt]
 & &  3& $1.87_{-0.15}^{+0.16}$  & $0.75_{-0.05}^{+0.04}$    & $1.95_{-0.16}^{+0.20}$& $2.61_{-0.33}^{+0.35}$    & $3.57_{-0.38}^{+0.35}$ & 0.90& 357& \ 2.74& 30.81\\[2pt]
 & &  4& $1.69_{-0.16}^{+0.18}$  & $0.76_{-0.05}^{+0.04}$    & $2.06_{-0.24}^{+0.31}$& $2.42_{-0.31}^{+0.33}$    & $2.35_{-0.33}^{+0.35}$ & 0.98& 270& \ 2.11& 30.70\\[2pt]
 & &  5& $2.03_{-0.53}^{+1.16}$  & $0.61_{-0.22}^{+0.11}$    & $1.69_{-0.33}^{+0.61}$& $2.56_{-0.80}^{+2.28}$    & $2.09_{-0.56}^{+0.75}$ & 1.04&  57& \ 1.98& 30.67\\
\noalign{\medskip}\hline\noalign{\medskip}
17-066& JH 108& 
      1& $2.07_{-0.35}^{+0.40}$  & $1.29_{-0.37}^{+0.72}$    & $3.94_{-3.60}^{+10.04}$& $0.89_{-0.63}^{+1.79}$   & $1.81_{-1.34}^{+0.73}$ & 0.84&  93& \ 1.42& 30.52\\[2pt]
 & &  2& $2.12_{-0.29}^{+0.38}$  & $0.86_{-0.15}^{+0.17}$    & $2.14_{-0.31}^{+0.57}$& $0.78_{-0.33}^{+0.45}$    & $2.63_{-0.54}^{+0.52}$ & 1.12& 120& \ 1.57& 30.57\\[2pt]
 & &  3& $2.09_{-0.38}^{+0.46}$  & $0.73_{-0.08}^{+0.08}$    & $2.43_{-0.62}^{+1.36}$& $0.59_{-0.14}^{+0.12}$    & $0.89_{-0.26}^{+0.09}$ & 1.23&  97& \ 0.64& 30.18\\
\noalign{\medskip}\hline\noalign{\medskip}
21-039& HD 283572& 
      1& $0.66_{-0.10}^{+0.11}$  & $0.75_{-0.03}^{+0.03}$    & $1.94_{-0.20}^{+0.22}$& $5.48_{-1.06}^{+1.29}$    & $6.37_{-0.68}^{+0.63}$ & 0.99& 347& \ 5.46& 31.11\\[2pt]
 & &  2& $0.72_{-0.07}^{+0.05}$  & $0.74_{-0.02}^{+0.01}$    & $2.03_{-0.10}^{+0.07}$& $4.61_{-0.66}^{+0.40}$    & $8.23_{-0.24}^{+0.38}$ & 1.05& 825& \ 6.05& 31.15\\[2pt]
 & &  3& $0.59_{-0.02}^{+0.04}$  & $0.76_{-0.01}^{+0.01}$    & $2.22_{-0.12}^{+0.08}$& $3.34_{-0.45}^{+0.33}$    & $8.96_{-0.19}^{+0.35}$ & 1.09& 782& \ 6.21& 31.16\\[2pt]
 & &  4& $0.64_{-0.19}^{+0.18}$  & $0.70_{-0.06}^{+0.05}$    & $1.83_{-0.16}^{+0.19}$& $3.86_{-1.15}^{+1.25}$    & $7.50_{-0.66}^{+0.66}$ & 0.99& 220& \ 5.39& 31.10\\[2pt]
 & &  5& $0.68_{-0.05}^{+0.06}$  & $0.75_{-0.01}^{+0.02}$    & $1.87_{-0.08}^{+0.07}$& $4.14_{-0.38}^{+0.56}$    & $6.23_{-0.28}^{+0.26}$ & 1.05& 715& \ 4.91& 31.06\\[2pt]
 & &  6& $0.64_{-0.06}^{+0.06}$  & $0.73_{-0.03}^{+0.02}$    & $1.67_{-0.08}^{+0.06}$& $3.48_{-0.42}^{+0.47}$    & $6.11_{-0.24}^{+0.40}$ & 0.98& 616& \ 4.50& 31.02\\[2pt]
 & &  7& $0.66_{-0.07}^{+0.07}$  & $0.74_{-0.02}^{+0.02}$    & $1.86_{-0.08}^{+0.09}$& $4.12_{-0.66}^{+0.59}$    & $5.97_{-0.31}^{+0.38}$ & 1.16& 553& \ 4.74& 31.05\\
\noalign{\medskip}\hline\noalign{\medskip}
22-047& XZ Tau& 
      1& $2.45_{-0.44}^{+0.61}$  & $0.73_{-0.11}^{+0.08}$    & $3.44_{-0.95}^{+2.31}$& $0.45_{-0.09}^{+0.12}$    & $0.31_{-0.07}^{+0.09}$ & 1.44&  75& \ 0.36& 29.93\\[2pt]
 & &  2& $2.05_{-0.39}^{+0.47}$  & $0.82_{-0.09}^{+0.15}$    & $5.72_{-1.71}^{+4.49}$& $0.49_{-0.12}^{+0.16}$    & $0.68_{-0.09}^{+0.12}$ & 0.90&  80& \ 0.65& 30.18\\[2pt]
 & &  3& $1.91_{-0.33}^{+0.39}$  & $0.77_{-0.12}^{+0.13}$    & $6.03_{-1.51}^{+2.06}$& $0.45_{-0.14}^{+0.14}$    & $1.25_{-0.09}^{+0.16}$ & 1.04& 143& \ 1.00& 30.37\\[2pt]
 & &  4& $2.22_{-0.21}^{+0.23}$  & $0.78_{-0.09}^{+0.13}$    & $4.42_{-0.45}^{+0.64}$& $0.49_{-0.12}^{+0.12}$    & $1.83_{-0.12}^{+0.12}$ & 0.95& 278& \ 1.30& 30.49\\[2pt]
 & &  5& $2.56_{-0.26}^{+0.29}$  & $0.79_{-0.09}^{+0.09}$    & $3.50_{-0.42}^{+0.52}$& $0.82_{-0.19}^{+0.19}$    & $2.23_{-0.16}^{+0.21}$ & 0.94& 230& \ 1.60& 30.58\\
\noalign{\medskip}\hline\noalign{\medskip}
22-089& L1551 51& 
      1& $0.87_{-0.24}^{+0.28}$  & $0.65_{-0.17}^{+0.14}$    & $1.50_{-0.23}^{+0.49}$& $1.01_{-0.26}^{+0.31}$    & $1.53_{-0.47}^{+0.24}$ & 0.92&  63& \ 1.08& 30.40\\[2pt]
 & &  2& $1.00_{-0.21}^{+0.69}$  & $0.62_{-0.15}^{+0.06}$    & $1.25_{-0.21}^{+0.59}$& $1.13_{-0.26}^{+0.31}$    & $0.61_{-0.24}^{+0.21}$ & 1.19&  90& \ 0.72& 30.23\\[2pt]
 & &  3& $0.62_{-0.18}^{+0.21}$  & $0.63_{-0.06}^{+0.05}$    & $1.37_{-0.23}^{+0.63}$& $0.85_{-0.16}^{+0.12}$    & $0.38_{-0.12}^{+0.14}$ & 0.95& 101& \ 0.59& 30.14\\
\clearpage
23-032+24-028& V410 Tau& 
      1& $0.10_{\phantom{-0.00}}$& $0.74_{-0.02}^{+0.02}$    & $1.70_{-0.08}^{+0.14}$& $1.79_{-0.14}^{+0.12}$    & $2.00_{-0.14}^{+0.14}$ & 1.27& 685& \ 1.63& 30.58\\[2pt]
 & &  2& $0.11_{-0.05}^{+0.05}$  & $0.75_{-0.03}^{+0.03}$    & $2.05_{-0.15}^{+0.25}$& $1.90_{-0.16}^{+0.14}$    & $2.80_{-0.26}^{+0.19}$ & 1.07& 540& \ 2.10& 30.69\\[2pt]
 & &  3& $0.09_{-0.06}^{+0.05}$  & $0.75_{-0.03}^{+0.04}$    & $2.30_{-0.26}^{+0.22}$& $1.90_{-0.12}^{+0.16}$    & $3.39_{-0.24}^{+0.31}$ & 1.18& 498& \ 2.43& 30.76\\[2pt]
 & &  4& $0.15_{-0.05}^{+0.05}$  & $0.79_{-0.02}^{+0.03}$    & $2.48_{-0.20}^{+0.23}$& $2.52_{-0.19}^{+0.19}$    & $3.60_{-0.26}^{+0.26}$ & 1.06& 494& \ 2.84& 30.82\\[2pt]
 & &  5& $0.19_{-0.07}^{+0.07}$  & $0.77_{-0.03}^{+0.04}$    & $2.33_{-0.26}^{+0.29}$& $3.01_{-0.28}^{+0.31}$    & $4.04_{-0.38}^{+0.42}$ & 1.08& 373& \ 3.25& 30.88\\[2pt]
 & &  6& $0.25_{-0.04}^{+0.05}$  & $0.75_{-0.02}^{+0.03}$    & $2.10_{-0.12}^{+0.26}$& $3.53_{-0.24}^{+0.31}$    & $4.56_{-0.42}^{+0.33}$ & 1.02& 571& \ 3.63& 30.93\\[2pt]
 & &  7& $0.29_{-0.14}^{+0.15}$  & $0.82_{-0.05}^{+0.09}$    & $3.67_{-0.92}^{+1.66}$& $4.47_{-0.66}^{+0.63}$    & $4.42_{-0.78}^{+0.87}$ & 1.00& 144& \ 4.41& 31.02\\[2pt]
 & &  8& $0.33_{-0.08}^{+0.09}$  & $0.77_{-0.04}^{+0.04}$    & $2.50_{-0.25}^{+0.29}$& $3.83_{-0.42}^{+0.42}$    & $6.70_{-0.59}^{+0.59}$ & 0.87& 355& \ 4.95& 31.07\\[2pt]
 & &  9& $0.35_{-0.07}^{+0.08}$  & $0.78_{-0.04}^{+0.04}$    & $2.02_{-0.16}^{+0.25}$& $3.25_{-0.38}^{+0.40}$    & $5.67_{-0.56}^{+0.52}$ & 1.04& 369& \ 3.99& 30.97\\[2pt]
 & & 10& $0.37_{-0.08}^{+0.09}$  & $0.77_{-0.04}^{+0.04}$    & $2.07_{-0.21}^{+0.43}$& $3.27_{-0.40}^{+0.45}$    & $4.35_{-0.63}^{+0.49}$ & 0.88& 292& \ 3.40& 30.90\\[2pt]
 & & 11& $0.21_{-0.06}^{+0.07}$  & $0.76_{-0.04}^{+0.03}$    & $1.90_{-0.17}^{+0.21}$& $3.03_{-0.28}^{+0.28}$    & $3.22_{-0.35}^{+0.35}$ & 1.15& 377& \ 4.27& 31.00\\[2pt]
 & & 12& $0.20_{-0.07}^{+0.09}$  & $0.77_{-0.04}^{+0.04}$    & $1.99_{-0.23}^{+0.34}$& $2.66_{-0.31}^{+0.31}$    & $2.73_{-0.40}^{+0.38}$ & 1.00& 291& \ 2.38& 30.75\\[2pt]
 & & 13& $0.14_{-0.03}^{+0.03}$  & $0.73_{-0.01}^{+0.02}$    & $1.79_{-0.10}^{+0.11}$& $1.98_{-0.09}^{+0.12}$    & $2.52_{-0.14}^{+0.07}$ & 1.15& 818& \ 1.94& 30.66\\[2pt]
 & & 14& $0.07_{-0.06}^{+0.06}$  & $0.77_{-0.06}^{+0.03}$    & $1.75_{-0.20}^{+0.16}$& $1.55_{-0.28}^{+0.16}$    & $2.09_{-0.19}^{+0.35}$ & 1.03& 448& \ 1.58& 30.57\\[2pt]
 & & 15& $0.12_{-0.10}^{+0.10}$  & $0.79_{-0.05}^{+0.05}$    & $2.49_{-0.37}^{+0.54}$& $1.74_{-0.24}^{+0.28}$    & $2.66_{-0.35}^{+0.35}$ & 1.03& 212& \ 2.06& 30.69\\[2pt]
 & & 16& $0.10_{\phantom{-0.00}}$& $0.58_{-0.17}^{+0.12}$    & $1.53_{-0.20}^{+0.18}$& $0.99_{-0.28}^{+0.28}$    & $2.78_{-0.35}^{+0.38}$ & 1.19&  93& \ 1.58& 30.57\\[2pt]
 & & 17& $0.10_{\phantom{-0.00}}$& $0.76_{-0.04}^{+0.04}$    & $1.81_{-0.16}^{+0.32}$& $1.41_{-0.16}^{+0.19}$    & $1.81_{-0.24}^{+0.21}$ & 0.91& 311& \ 1.41& 30.52\\
\noalign{\medskip}\hline\noalign{\medskip}
23-047+24-040& V892 Tau& 
      1& $8.96_{-1.70}^{+1.67}$  & $0.82_{-0.13}^{+0.22}$    & $2.98_{-0.47}^{+0.84}$& $3.15_{-1.79}^{+2.78}$    & $3.39_{-0.73}^{+0.63}$ & 1.10& 254& \ 3.13& 30.87\\[2pt]
 & &  2& $10.51_{-3.72}^{+1.44}$ & $0.69_{-0.17}^{+0.38}$    & $2.93_{-0.62}^{+1.73}$& $5.86_{-2.94}^{+7.31}$    & $4.99_{-1.67}^{+1.25}$ & 0.79& 77 & \ 5.08& 31.08\\[2pt]
 & &  3& $8.90_{-0.77}^{+1.10}$  & $0.92_{-0.16}^{+0.11}$    & $2.87_{-0.36}^{+0.52}$& $3.25_{-1.08}^{+1.27}$    & $4.00_{-0.73}^{+0.75}$ & 0.97& 338& \ 3.46& 30.91\\[2pt]
 & &  4& $8.14_{-0.92}^{+1.13}$  & $1.06_{-0.18}^{+0.18}$    & $3.02_{-0.41}^{+0.70}$& $2.89_{-1.39}^{+1.86}$    & $4.99_{-1.18}^{+1.03}$ & 1.05& 194& \ 3.86& 30.96\\[2pt]
 & &  5& $9.59_{-1.06}^{+1.20}$  & $0.81_{-0.12}^{+0.15}$    & $2.37_{-0.27}^{+0.40}$& $4.59_{-1.83}^{+2.54}$    & $4.19_{-0.80}^{+0.73}$ & 1.13& 242& \ 3.99& 30.97\\[2pt]
 & &  6& $12.33_{-1.59}^{+1.66}$ & $0.38_{-0.05}^{+0.07}$    & $1.84_{-0.16}^{+0.18}$& $13.90_{-6.87}^{+12.30}$  & $5.13_{-0.59}^{+0.66}$ & 0.97& 232& \ 7.52& 31.25\\[2pt]
 & &  7& $11.05_{-1.85}^{+1.24}$ & $0.49_{-0.08}^{+0.19}$    & $1.95_{-0.20}^{+0.42}$& $6.00_{-3.41}^{+3.53}$    & $3.69_{-0.92}^{+0.61}$ & 1.05& 295& \ 4.09& 30.98\\[2pt]
 & &  8& $10.73_{-0.67}^{+0.77}$ & $0.49_{\phantom{-0.00}}$  & $5.71_{-1.32}^{+2.04}$& $6.00_{\phantom{-0.00}}$  & $6.21_{-0.61}^{+0.75}$ & 1.03& 90 & \ 6.38& 31.18\\[2pt]
 & &  9& $12.07_{-1.05}^{+1.20}$ & $0.49_{\phantom{-0.00}}$  & $8.04_{-2.21}^{+4.06}$& $6.00_{\phantom{-0.00}}$  & $13.71_{-1.15}^{+1.58}$& 1.02& 85 & 11.60 & 31.44\\[2pt]
 & & 10& $11.68_{-0.86}^{+0.98}$ & $0.49_{\phantom{-0.00}}$  & $8.62_{-1.94}^{+4.12}$& $6.00_{\phantom{-0.00}}$  & $18.04_{-1.15}^{+1.48}$& 0.72& 124& 14.59 & 31.54\\[2pt]
 & & 11& $11.53_{-0.78}^{+0.79}$ & $0.49_{\phantom{-0.00}}$  & $8.23_{-1.59}^{+2.99}$& $6.00_{\phantom{-0.00}}$  & $21.38_{-1.34}^{+1.48}$& 0.99& 181& 16.72 & 31.59\\[2pt]
 & & 12& $11.26_{-0.26}^{+0.28}$ & $0.49_{\phantom{-0.00}}$  & $7.24_{-0.59}^{+0.65}$& $6.00_{\phantom{-0.00}}$  & $26.72_{-0.66}^{+0.73}$& 0.99& 862& 20.08 & 31.67\\
\noalign{\medskip}\hline\noalign{\medskip}
23-050+24-042& V410 X7& 
      1& $7.41_{-1.91}^{+5.79}$  & $0.77_{-0.53}^{+0.37}$    & $2.07_{-0.39}^{+2.38}$& $0.33_{-0.24}^{+1.60}$    & $0.35_{-0.19}^{+0.09}$ & 1.16& 130& \ 0.30& 29.85\\[2pt]
 & &  2& $8.19_{-1.04}^{+1.38}$  & $1.36_{-0.36}^{+0.68}$    & $5.43_{-1.47}^{+2.36}$& $2.42_{-1.76}^{+1.58}$    & $1.69_{-1.08}^{+1.76}$ & 0.92& 105& \ 2.07& 30.69\\[2pt]
 & &  3& $8.34_{-1.67}^{+1.41}$  & $0.68_{-0.44}^{+0.28}$    & $2.55_{-0.47}^{+0.49}$& $0.63_{-0.52}^{+1.51}$    & $2.07_{-0.31}^{+0.56}$ & 0.97& 139& \ 1.29& 30.48\\[2pt]
 & &  4& $10.46_{-1.87}^{+4.81}$ & $0.68_{-0.44}^{+0.24}$    & $1.78_{-0.44}^{+1.44}$& $2.02_{-1.32}^{+2.09}$    & $1.34_{-0.73}^{+1.69}$ & 1.06& 116& \ 1.45& 30.53\\
\noalign{\medskip}\hline\noalign{\medskip}
23-056+24-047& Hubble 4& 
      1& $2.32_{-0.09}^{+0.09}$  & $0.74_{-0.02}^{+0.02}$    & $1.81_{-0.15}^{+0.12}$& $2.54_{-0.16}^{+0.16}$    & $2.49_{-0.19}^{+0.24}$ & 1.19& 680& \ 2.18& 30.71\\[2pt]
 & &  2& $1.58_{-0.32}^{+0.37}$  & $0.77_{-0.10}^{+0.07}$    & $2.28_{-0.45}^{+0.76}$& $1.53_{-0.35}^{+0.38}$    & $1.76_{-0.38}^{+0.40}$ & 0.96& 89 & \ 1.50& 30.55\\[2pt]
 & &  3& $2.15_{-0.19}^{+0.20}$  & $0.76_{-0.05}^{+0.04}$    & $1.84_{-0.20}^{+0.24}$& $2.07_{-0.26}^{+0.31}$    & $2.07_{-0.28}^{+0.33}$ & 1.00& 239& \ 1.80& 30.63\\[2pt]
 & &  4& $2.24_{-0.13}^{+0.13}$  & $0.77_{-0.04}^{+0.03}$    & $2.08_{-0.15}^{+0.28}$& $2.30_{-0.24}^{+0.26}$    & $2.80_{-0.31}^{+0.24}$ & 1.00& 443& \ 2.28& 30.73\\[2pt]
 & &  5& $2.09_{-0.16}^{+0.18}$  & $0.76_{-0.04}^{+0.04}$    & $1.87_{-0.21}^{+0.22}$& $2.12_{-0.24}^{+0.26}$    & $2.14_{-0.28}^{+0.28}$ & 1.06& 302& \ 1.85& 30.64\\[2pt]
 & &  6& $2.15_{-0.13}^{+0.14}$  & $0.73_{-0.12}^{+0.04}$    & $1.63_{-0.13}^{+0.36}$& $1.86_{-0.21}^{+0.31}$    & $1.88_{-0.33}^{+0.35}$ & 1.16& 494& \ 1.60& 30.58\\[2pt]
 & &  7& $2.29_{-0.28}^{+0.44}$  & $0.73_{-0.10}^{+0.05}$    & $2.63_{-0.58}^{+1.15}$& $2.47_{-0.38}^{+0.42}$    & $1.79_{-0.38}^{+0.42}$ & 0.92& 165& \ 1.94& 30.66\\
\noalign{\medskip}\hline\noalign{\medskip}
23-074+24-061& V819 Tau& 
      1& $2.11_{-0.20}^{+0.23}$  & $0.40_{-0.02}^{+0.02}$    & $1.31_{-0.07}^{+0.13}$& $1.41_{-0.19}^{+0.26}$    & $0.92_{-0.07}^{+0.07}$ & 1.06& 438& \ 0.93& 30.34\\[2pt]
 & &  2& $2.24_{-0.51}^{+0.58}$  & $0.38_{-0.06}^{+0.09}$    & $1.66_{-0.19}^{+0.49}$& $1.60_{-0.61}^{+0.85}$    & $2.38_{-0.35}^{+0.28}$ & 0.87& 80 & \ 1.63& 30.58\\[2pt]
 & &  3& $1.72_{-0.42}^{+0.49}$  & $0.37_{-0.05}^{+0.06}$    & $1.22_{-0.11}^{+0.12}$& $1.08_{-0.35}^{+0.49}$    & $1.18_{-0.16}^{+0.14}$ & 0.89& 100& \ 0.89& 30.32\\[2pt]
 & &  4& $2.23_{-0.34}^{+0.46}$  & $0.36_{-0.04}^{+0.03}$    & $1.20_{-0.12}^{+0.12}$& $1.34_{-0.31}^{+0.47}$    & $0.89_{-0.12}^{+0.12}$ & 0.93& 218& \ 0.87& 30.31\\
\clearpage
26-072& HBC 427& 
      1& $0.20_{-0.19}^{+0.20}$  & $0.79_{-0.07}^{+0.06}$    & $2.14_{-0.43}^{+0.90}$& $1.51_{-0.28}^{+0.26}$    & $1.48_{-0.33}^{+0.35}$ & 0.88& 104& \ 1.34& 30.50\\[2pt]
 & &  2& $0.29_{-0.28}^{+0.32}$  & $0.93_{-0.18}^{+0.16}$    & $6.18_{-2.05}^{+5.28}$& $2.00_{-0.82}^{+0.92}$    & $5.39_{-0.82}^{+0.94}$ & 0.99& 73 & \ 4.32& 31.01\\[2pt]
 & &  3& $0.57_{-0.23}^{+0.28}$  & $0.50_{-0.14}^{+0.17}$    & $3.31_{-0.44}^{+0.75}$& $1.81_{-0.73}^{+0.92}$    & $10.89_{-1.01}^{+0.94}$& 0.95& 132& \ 6.70& 31.20\\[2pt]
 & &  4& $0.49_{-0.26}^{+0.29}$  & $0.89_{-0.13}^{+0.17}$    & $2.88_{-0.64}^{+1.35}$& $2.68_{-1.01}^{+1.36}$    & $6.68_{-1.62}^{+1.39}$ & 0.94& 74 & \ 4.62& 31.04\\[2pt]
 & &  5& $0.44_{-0.13}^{+0.15}$  & $0.80_{-0.06}^{+0.08}$    & $2.43_{-0.29}^{+0.35}$& $2.16_{-0.38}^{+0.40}$    & $5.10_{-0.52}^{+0.52}$ & 1.01& 190& \ 3.41& 30.90\\[2pt]
 & &  6& $0.51_{-0.12}^{+0.12}$  & $0.74_{-0.05}^{+0.04}$    & $2.01_{-0.20}^{+0.38}$& $2.23_{-0.31}^{+0.31}$    & $3.27_{-0.45}^{+0.38}$ & 1.31& 220& \ 2.45& 30.76\\[2pt]
 & &  7& $0.28_{-0.16}^{+0.17}$  & $0.83_{-0.06}^{+0.15}$    & $1.89_{-0.43}^{+0.80}$& $2.23_{-0.45}^{+0.49}$    & $1.83_{-1.15}^{+0.63}$ & 1.10& 128& \ 1.82& 30.63\\[2pt]
 & &  8& $0.32_{-0.10}^{+0.12}$  & $0.74_{-0.06}^{+0.04}$    & $1.54_{-0.14}^{+0.20}$& $1.76_{-0.28}^{+0.28}$    & $1.90_{-0.31}^{+0.33}$ & 1.15& 217& \ 1.57& 30.57\\
\noalign{\medskip}\hline\noalign{\medskip}
28-100& BP~Tau& 
      1& $0.96_{-0.22}^{+0.24}$  & $0.37_{-0.03}^{+0.04}$    & $1.57_{-0.11}^{+0.11}$& $0.63_{-0.12}^{+0.14}$    & $0.94_{-0.07}^{+0.05}$ & 1.25& 226& \ 0.63& 30.17\\[2pt]
 & &  2& $1.00_{\phantom{-0.00}}$& $0.41_{-0.06}^{+0.11}$    & $2.43_{-0.44}^{+0.75}$& $0.68_{-0.12}^{+0.14}$    & $1.25_{-0.19}^{+0.16}$ & 0.76& 75 & \ 0.87& 30.31\\[2pt]
 & &  3& $1.00_{\phantom{-0.00}}$& $0.79_{-0.23}^{+0.24}$    & $4.04_{-1.05}^{+1.95}$& $0.78_{-0.24}^{+0.26}$    & $1.93_{-0.31}^{+0.33}$ & 0.98& 79 & \ 1.46& 30.54\\[2pt]
 & &  4& $1.26_{-0.13}^{+0.12}$  & $0.75_{-0.06}^{+0.06}$    & $2.65_{-0.17}^{+0.23}$& $0.89_{-0.16}^{+0.14}$    & $3.29_{-0.21}^{+0.21}$ & 0.93& 419& \ 2.06& 30.69\\[2pt]
 & &  5& $1.30_{-0.20}^{+0.29}$  & $0.79_{-0.09}^{+0.06}$    & $2.39_{-0.54}^{+0.49}$& $1.55_{-0.26}^{+0.28}$    & $1.93_{-0.33}^{+0.52}$ & 1.23& 133& \ 1.59& 30.57\\[2pt]
 & &  6& $0.85_{-0.27}^{+0.94}$  & $0.77_{-0.09}^{+0.08}$    & $2.13_{-0.30}^{+0.47}$& $1.06_{-0.21}^{+0.24}$    & $1.55_{-0.28}^{+0.31}$ & 0.80& 101& \ 1.17& 30.44\\[2pt]
 & &  7& $0.95_{-0.19}^{+2.64}$  & $0.78_{-0.08}^{+0.06}$    & $1.82_{-0.19}^{+0.24}$& $0.80_{-0.14}^{+0.16}$    & $1.27_{-0.21}^{+0.19}$ & 1.14& 156& \ 0.90& 30.33\\[2pt]
 & &  8& $0.64_{-0.38}^{+0.43}$  & $0.78_{-0.08}^{+0.21}$    & $1.60_{-0.15}^{+0.57}$& $0.52_{-0.19}^{+0.16}$    & $0.94_{-0.26}^{+0.16}$ & 0.97& 103& \ 0.62& 30.16\\[2pt]
 & &  9& $1.27_{-0.20}^{+0.26}$  & $0.34_{-0.02}^{+0.02}$    & $1.50_{-0.08}^{+0.07}$& $0.66_{-0.12}^{+0.14}$    & $0.92_{-0.05}^{+0.05}$ & 0.99& 345& \ 0.63& 30.17\\[2pt]
 & & 10& $1.00_{\phantom{-0.00}}$& $0.40_{-0.03}^{+0.05}$    & $1.62_{-0.12}^{+0.25}$& $0.54_{-0.07}^{+0.05}$    & $0.94_{-0.09}^{+0.07}$ & 1.20& 179& \ 0.60& 30.15\\[2pt]
 & & 11& $0.87_{-0.17}^{+0.19}$  & $0.37_{-0.02}^{+0.03}$    & $1.51_{-0.07}^{+0.09}$& $0.49_{-0.07}^{+0.07}$    & $0.71_{-0.05}^{+0.05}$ & 1.05& 348& \ 0.48& 30.05\\[2pt]
 & & 12& $0.69_{-0.21}^{+0.26}$  & $0.38_{-0.03}^{+0.05}$    & $1.87_{-0.22}^{+0.21}$& $0.49_{-0.09}^{+0.14}$    & $1.01_{-0.07}^{+0.07}$ & 1.03& 218& \ 0.64& 30.18\\[2pt]
 & & 13& $1.03_{-0.40}^{+0.35}$  & $0.31_{-0.03}^{+0.07}$    & $1.29_{-0.07}^{+0.17}$& $0.45_{-0.16}^{+0.19}$    & $0.94_{-0.21}^{+0.07}$ & 0.98& 198& \ 0.56& 30.12\\
\noalign{\medskip}\hline\hline\noalign{\medskip}
05-031& HD 283810& 
      1& $0.01_{-0.01}^{+0.16}$  & $0.41_{-0.03}^{+0.05}$    & $0.98_{-0.07}^{+0.08}$& $2.16_{-0.33}^{+0.45}$    & $1.95_{-0.28}^{+0.28}$ & 1.21& 165& \ 1.66& 30.22\\[2pt]
 & &  2& $0.00_{-0.00}^{+0.02}$  & $0.41_{-0.02}^{+0.02}$    & $0.79_{-0.04}^{+0.05}$& $1.69_{-0.31}^{+0.28}$    & $1.93_{-0.26}^{+0.31}$ & 1.49& 420& \ 1.48& 30.17\\[2pt]
 & &  3& $0.01_{\phantom{-0.00}}$& $0.60_{-0.06}^{+0.06}$    & $1.56_{-0.19}^{+0.25}$& $3.01_{-0.49}^{+0.45}$    & $4.66_{-0.57}^{+0.59}$ & 1.07& 172& \ 3.24& 30.51\\[2pt]
 & &  4& $0.01_{\phantom{-0.00}}$& $0.69_{-0.04}^{+0.06}$    & $2.27_{-0.72}^{+3.24}$& $3.53_{-0.78}^{+0.54}$    & $1.72_{-0.68}^{+0.63}$ & 1.18& 88 & \ 2.32& 30.37\\[2pt]
 & &  5& $0.01_{\phantom{-0.00}}$& $0.19_{-0.08}^{+0.07}$    & $0.75_{-0.13}^{+0.05}$& $0.96_{-0.52}^{+14.89}$   & $4.00_{-0.47}^{+2.19}$ & 1.32& 106& \ 1.89& 30.28\\
\noalign{\medskip}\hline\noalign{\medskip}
16-031& J04195676+27& 
      4& $0.11_{-0.11}^{+0.14}$  & $\ldots_{\phantom{-0.00}}$& $1.85_{-0.25}^{+0.25}$& $\ldots_{\phantom{-0.00}}$& $3.34_{-0.21}^{+0.21}$ & 1.10& 110& \ 1.48& 30.54\\[2pt]
 & &  5& $0.10_{\phantom{-0.00}}$& $\ldots_{\phantom{-0.00}}$& $1.65_{-0.11}^{+0.17}$& $\ldots_{\phantom{-0.00}}$& $1.81_{-0.07}^{+0.09}$ & 1.29& 91 & \ 0.78& 30.26\\[2pt]
 & &  6& $0.10_{\phantom{-0.00}}$& $\ldots_{\phantom{-0.00}}$& $1.63_{-0.10}^{+0.21}$& $\ldots_{\phantom{-0.00}}$& $2.35_{-0.02}^{+0.19}$ & 1.50& 123& \ 1.74& 30.61\\[2pt]
 & &  7& $0.10_{\phantom{-0.00}}$& $\ldots_{\phantom{-0.00}}$& $3.63_{-0.64}^{+0.78}$& $\ldots_{\phantom{-0.00}}$& $6.56_{-0.33}^{+0.38}$ & 0.87& 69 & \ 3.71& 30.94\\[2pt]
 & &  8& $0.10_{\phantom{-0.00}}$& $\ldots_{\phantom{-0.00}}$& $2.38_{-0.37}^{+0.39}$& $\ldots_{\phantom{-0.00}}$& $9.88_{-0.21}^{+0.89}$ & 1.02& 68 & \ 4.80& 31.05\\[2pt]
 & &  9& $0.10_{\phantom{-0.00}}$& $\ldots_{\phantom{-0.00}}$& $3.02_{-0.33}^{+0.30}$& $\ldots_{\phantom{-0.00}}$& $16.20_{-0.02}^{+1.15}$& 1.12& 161& \ 8.58& 31.30\\[2pt]
 & & 10& $0.10_{\phantom{-0.00}}$& $\ldots_{\phantom{-0.00}}$& $2.01_{-0.25}^{+0.44}$& $\ldots_{\phantom{-0.00}}$& $13.33_{-0.73}^{+0.68}$& 1.05& 65 & \ 6.08& 31.16\\[2pt]
 & & 11& $0.10_{\phantom{-0.00}}$& $\ldots_{\phantom{-0.00}}$& $1.92_{-0.25}^{+0.24}$& $\ldots_{\phantom{-0.00}}$& $11.22_{-0.56}^{+0.56}$& 1.17& 78 & \ 5.05& 31.07\\[2pt]
 & & 12& $0.10_{\phantom{-0.00}}$& $\ldots_{\phantom{-0.00}}$& $1.63_{-0.12}^{+0.22}$& $\ldots_{\phantom{-0.00}}$& $7.06_{-0.09}^{+0.61}$ & 1.51& 102& \ 3.02& 30.85\\[2pt]
 & & 13& $0.10_{\phantom{-0.00}}$& $\ldots_{\phantom{-0.00}}$& $1.64_{-0.10}^{+0.17}$& $\ldots_{\phantom{-0.00}}$& $3.81_{-0.16}^{+0.16}$ & 1.26& 100& \ 1.63& 30.58\\[2pt]
 & & 14& $0.10_{\phantom{-0.00}}$& $\ldots_{\phantom{-0.00}}$& $1.24_{-0.06}^{+0.06}$& $\ldots_{\phantom{-0.00}}$& $1.58_{-0.09}^{+0.07}$ & 1.47& 95 & \ 0.65& 30.18\\
\noalign{\medskip}\hline\noalign{\medskip}
22-024& HD 285845& 
      1& $0.42_{-0.10}^{+0.10}$  & $0.64_{-0.04}^{+0.03}$    & $2.52_{-0.30}^{+0.54}$& $3.27_{-0.24}^{+0.31}$    & $3.27_{-0.38}^{+0.33}$ & 0.93& 295& \ 3.00& 30.66\\[2pt]
 & &  2& $0.33_{-0.21}^{+0.23}$  & $0.77_{-0.06}^{+0.06}$    & $3.49_{-1.28}^{+6.30}$& $4.09_{-0.71}^{+0.80}$    & $3.32_{-0.87}^{+1.01}$ & 0.96& 86 & \ 3.58& 30.74\\[2pt]
 & &  3& $0.47_{-0.12}^{+0.06}$  & $0.61_{-0.16}^{+0.04}$    & $2.86_{-0.44}^{+0.49}$& $3.50_{-0.38}^{+0.42}$    & $5.29_{-0.47}^{+0.63}$ & 1.09& 232& \ 4.21& 30.81\\[2pt]
 & &  4& $0.36_{-0.17}^{+0.19}$  & $0.66_{-0.06}^{+0.06}$    & $3.10_{-0.82}^{+1.43}$& $3.39_{-0.47}^{+0.54}$    & $3.50_{-0.59}^{+0.68}$ & 1.01& 119& \ 3.30& 30.70\\[2pt]
 & &  5& $0.40_{-0.11}^{+0.13}$  & $0.65_{-0.03}^{+0.09}$    & $3.01_{-0.65}^{+0.87}$& $3.20_{-0.28}^{+0.31}$    & $2.70_{-0.33}^{+0.47}$ & 1.09& 215& \ 2.78& 30.63\\[2pt]
 & &  6& $0.39_{-0.09}^{+0.09}$  & $0.65_{-0.03}^{+0.05}$    & $3.04_{-0.41}^{+0.48}$& $2.99_{-0.21}^{+0.28}$    & $3.62_{-0.31}^{+0.33}$ & 0.90& 339& \ 3.19& 30.69\\[2pt]
 & &  7& $0.51_{-0.08}^{+0.09}$  & $0.64_{-0.04}^{+0.03}$    & $1.65_{-0.10}^{+0.33}$& $2.70_{-0.24}^{+0.40}$    & $3.57_{-0.45}^{+0.31}$ & 0.91& 368& \ 2.67& 30.61\\[2pt]
 & &  8& $0.34_{-0.22}^{+0.22}$  & $0.67_{-0.05}^{+0.08}$    & $2.48_{-0.86}^{+3.41}$& $3.13_{-0.42}^{+0.45}$    & $1.34_{-0.59}^{+0.49}$ & 0.71& 105& \ 1.99& 30.48\\[2pt]
 & &  9& $0.32_{-0.12}^{+0.12}$  & $0.68_{-0.03}^{+0.03}$    & $1.82_{-0.27}^{+0.32}$& $2.73_{-0.24}^{+0.24}$    & $1.11_{-0.26}^{+0.24}$ & 0.95& 191& \ 1.64& 30.40\\[2pt]
 & & 10& $0.20_{-0.20}^{+0.20}$  & $0.66_{-0.58}^{+0.06}$    & $1.37_{-1.09}^{+0.39}$& $2.02_{-0.47}^{+0.35}$    & $1.11_{-0.31}^{+1.39}$ & 0.67& 65 & \ 1.32& 30.30\\[2pt]
 & & 11& $0.59_{-0.11}^{+0.09}$  & $0.66_{-0.06}^{+0.02}$    & $1.92_{-0.24}^{+0.29}$& $2.80_{-0.26}^{+0.31}$    & $2.16_{-0.21}^{+0.28}$ & 0.94& 217& \ 2.16& 30.52\\[2pt]
 & & 12& $0.38_{-0.11}^{+0.10}$  & $0.71_{-0.04}^{+0.03}$    & $1.84_{-0.37}^{+0.65}$& $2.61_{-0.35}^{+0.28}$    & $1.36_{-0.35}^{+0.40}$ & 1.04& 230& \ 1.72& 30.42\\[2pt]
 & & 13& $0.35_{-0.06}^{+0.05}$  & $0.66_{-0.02}^{+0.01}$    & $1.56_{-0.13}^{+0.14}$& $2.38_{-0.09}^{+0.09}$    & $0.89_{-0.09}^{+0.09}$ & 1.10& 428& \ 1.38& 30.32\\
\end{longtable}
}%

\end{document}